\documentclass[%
 reprint,
]{revtex4-2}

\usepackage{graphicx}
\usepackage{dcolumn}
\usepackage{amsmath}
\usepackage{xcolor}
\usepackage{bm}

\begin{document}


\title{Measuring the Neutron-Star Moment of Inertia using Nonlinear Orbital Variations from Lense-Thirring Precession}

\author{Emmanuel Fonseca}
    \email{emmanuel.fonseca@mail.wvu.edu}
    \affiliation{Department of Physics and Astronomy, West Virginia University, PO Box 6315, Morgantown, WV 26506, USA}
    \affiliation{Center for Gravitational Waves and Cosmology, West Virginia University, Chestnut Ridge Research Building, Morgantown, WV 26505, USA}

\date{\today}

\begin{abstract}
A future measurement of Lense-Thirring (LT) precession using a binary radio pulsar is expected to yield the pulsar's moment of inertia ($I_{\rm p}$). However, most of the known pulsar-binary systems expected to provide this opportunity will exhibit linear variations in the orbital elements, due to LT precession, that are difficult to separate from distance-dependent variations induced by other mechanisms. In this work, we demonstrate that the pulsar-timing signature of LT precession for an arbitrary orbital orientation produces nonlinear orbital variations after a sufficient amount of time; if detected, these nonlinear variations provide a means to constrain $I_{\rm p}$ without the need for independent measurements of orbital geometry or distance. We show through simulations that these nonlinear signatures are indeed detectable in pulsar-binary systems with an appropriate combination of timing precision and orbital geometry. Our simulations also show that nonlinear orbital variations from LT precession are expected to be detectable in PSR J1757$-$1854 and that its $I_{\rm p}$ can be constrained to precision as high as 10\% by 2040 even without knowledge of distance.
\end{abstract}

\maketitle

\section{Introduction}
\label{sec:intro}

A gyroscope undergoing relativistic two-body motion will exhibit two forms of evolution in its spin-orbit orientation. The dominant variation is de Sitter (or ``geodetic") precession, arising from parallel transport of the spin angular momentum vector (${\bf S}$) in curved spacetime induced by the orbital companion object \cite{bai88}. The second variation is Lense-Thirring (LT) precession of the orbital angular momentum vector ${\bf L}$ about the total angular momentum vector ${\bf J} = {\bf L} + {\bf S}$ due to the spin of the gyroscope \cite{ds88}. Radio pulsars -- rotating neutron stars emitting beamed radiation at radio frequencies -- serve as point-like gyroscopes, as their high masses ($m_{\rm p} \sim 1\textrm{ M}_\odot$) and small radii ($R_{\rm p} \sim 10\textrm{ km}$) correspond to large moments of inertia $I_{\rm p} \propto m_{\rm p}R_{\rm p}^2 \sim 10^{45}$ gm cm$^2$. Long-term observations of pulsars in compact, ``double-neutron-star'' (DNS) binary systems therefore provide the best avenue for resolving phenomena predicted by general relativity (GR), including LT precession, through gravitational interactions of point-like objects.

Recent works have shown that LT precession will arise in PSR J0737$-$3039A as a secular, post-Newtonian (PN) variation of its orbit that contributes to apsidal motion over time \cite{lbk+04,bs05,hf24}. The successful measurement of this linear variation in J0737$-$3039A will provide a new test of GR or, by assuming the correctness of GR, a test of extreme-density nuclear physics through a direct measurement of $I_{\rm p}$ \citep[e.g.,][]{lp16}. However, either effort requires the decomposition of ``observed" PN phenomena into those arising from GR and others arising from non-relativistic mechanisms. A common example of such sources is the relative acceleration between the observer and pulsar in the gravitational potential of the Milky Way galaxy, which induces variations in several orbital elements whose magnitude depend on distance \citep[e.g.,][]{mcm17}. The measurement of $I_{\rm p}$ from linear LT variations can therefore be subject to systematic uncertainties from modeling assumptions that go beyond phenomena predicted by GR.

However, the sense of both ${\bf L}$ and the pulsar's spin angular momentum vector ${\bf S}_{\rm p}$ evolve over time and in a periodic manner \footnote{In what follows, we assume that the spin of the pulsar's unseen companion neutron star (${\bf S}_{\rm c}$) is negligible, i.e., $|{\bf S_{\rm c}|} \ll |{\bf S_{\rm p}}|$ so that ${\bf J} \approx {\bf L} + {\bf S_{\rm p}}$. This assumption is made based on expectations from binary stellar evolution, where the progenitor of the second-formed neutron star has donated mass onto the pulsar and therefore lost angular momentum prior to its supernova explosion. LT precession in DNS systems will therefore most likely be dominated by the spin of the observed pulsar.}. Therefore, LT precession can induce nonlinear variations in the orbital geometry if a sufficient amount of precession has occurred. Nonlinear variations from LT precession, if detected, could serve to constrain $I_{\rm p}$ while circumventing sources of bias that affect only the linear variations. This nonlinearity in LT precession was previously noted in the context of pulsars orbiting black holes that possess substantial spin \citep{wk99}.

In this work, we explore the efficacy in detecting nonlinear variations from LT precession through pulsar timing. We show that relativistic DNS systems can exhibit nonlinearity in orbital evolution due to LT precession over $O(10\textrm{ yr})$ timescales, allowing for a timing-only derivation of $I_{\rm p}$ and the relevant spin-orbit orientation. We then conduct simulations to determine if any known DNS pulsars will exhibit nonlinear LT variations, finding that PSR J1757$-$1854 will likely do so by 2040.

\section{Variations from LT Precession}
\label{sec:vars}

\begin{figure}
    \centering
    \includegraphics[width=0.9\linewidth]{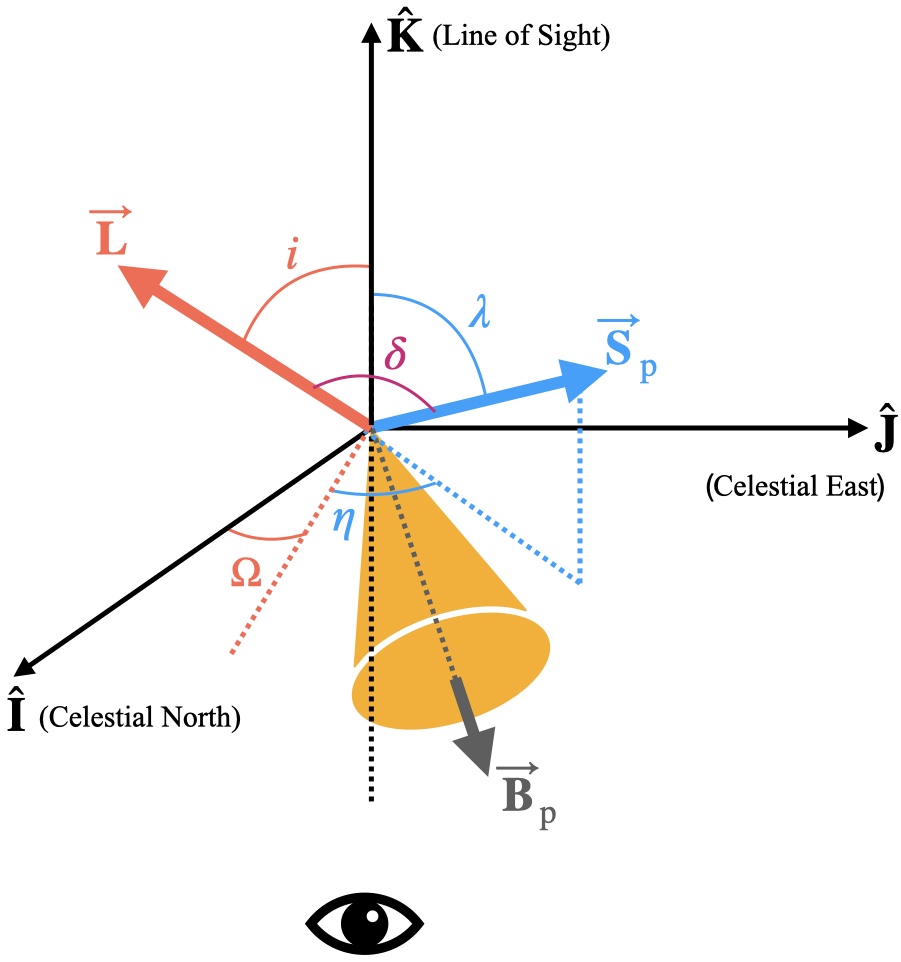}
    \caption{Orientation angles of ${\bf L}$ and ${\bf S}_{\rm p}$ for a binary pulsar that are mentioned in the text. Also shown is the pulsar's light cone (yellow) centered on its dipolar magnetic field (${\bf B}_{\rm p}$).}
    \label{fig:angles}
\end{figure}

Variations of the measurable Keplerian elements consist of terms intrinsic and extrinsic to the binary system \citep[e.g.,][]{dt92}. For example, pulse time-of-flight delays across the orbit are proportional to the semi-major axis of the pulsar's orbit ($a_{\rm p}$) projected onto an observer's line of sight, typically written as $x_{\rm int} = a_{\rm p}\sin i/c$ where $i$ is the orbital inclination angle and $c$ is the speed of light. The angles of orientation and their relationship to ${\bf L}$ and ${\bf S_{\rm p}}$ are shown in Figure \ref{fig:angles}. In addition to inclination, the {\it observed} amplitude ($x$) is further modulated by two factors: a factor $D$ quantifying the Doppler shift between an observer and the center of mass of a moving binary system; and a factor $\epsilon_{\rm p}$ arising from special-relativistic aberration of the pulsed radio signal. The observed $x \equiv x_{\rm obs} = D^{-1}(1+\epsilon_{\rm p})x_{\rm int}$. For the case of $x$, a variation in any of the relevant elements and/or ($D$, $\epsilon_{\rm p}$) produces a variation that can be written as

\begin{equation}
    \bigg(\frac{\dot{x}}{x}\bigg)_{\rm obs} = -\bigg(\frac{\dot{D}}{D}\bigg) + \bigg(\frac{\dot{a}_{\rm p}}{a_{\rm p}}\bigg) + \cot{i}\bigg(\frac{di}{dt}\bigg) + \frac{d\epsilon_{\rm p}}{dt}.
    \label{eq:xdot}
\end{equation}

The first and second terms of Equation \ref{eq:xdot} are known to arise from relative acceleration in the Milky Way Galaxy \citep[e.g.,][]{dt91} and PK energy loss due to gravitational radiation \citep{pet64}, respectively. The third term arises from an evolving orientation of the binary system, which can often be due to significant proper motion \cite{kop96}. We treat these contributions as constant since they vary on timescales much larger than those of existing data sets. By contrast, LT precession of ${\bf L}$ about ${\bf J}$ is expected to yield a PK variation in $i$ which, in accordance with GR, corresponds to a time-averaged rate

\begin{equation}
    \bigg(\frac{di}{dt}\bigg)_{\rm LT}^{\rm GR} = \frac{2\pi{G}}{P_{\rm s}c^2a^3(1-e^2)^{3/2}}\bigg(2 + \frac{3}{2}\frac{m_{\rm c}}{m_{\rm p}}\bigg)I_{\rm p}\sin\lambda\cos\eta,
    \label{eq:didtLT}
\end{equation}

\noindent where $G$ is Newton's gravitational constant, $P_{\rm s}$ is the pulsar's spin period, $a$ is the semi-major axis of the relative orbit, $e$ is the orbital eccentricity, $m_{\rm c}$ is the mass of the neutron-star companion with negligible spin, and ($\lambda$, $\eta$) are the polar angles that define the sense of ${\bf S}_{\rm p}$ relative to the plane of the sky, also shown in Figure 1 \cite{ds88}.

The fourth term arises from the varying aberration and is given by \cite{dt92},

\begin{equation}
    \frac{d\epsilon_{\rm p}}{dt} = -\frac{P_{\rm s}}{P_{\rm b}}\frac{\Omega_{\rm p}}{\sin^2\lambda(1-e^2)^{1/2}}\big(\cos\lambda\sin2\eta+\cot{i}\sin\lambda\cos\eta\big),
    \label{eq:aber}
\end{equation}

\noindent where $\Omega_{\rm p}$ is the rate of geodetic precession of ${\bf S}_{\rm p}$ about ${\bf J}$. According to GR \cite{bo75},

\begin{equation}
    \Omega_{\rm p} = \frac{1}{2}\frac{G^{2/3}}{c^2}\bigg(\frac{2\pi}{P_{\rm b}}\bigg)^{5/3}\frac{m_{\rm c}(4m_{\rm p} + 3m_{\rm c})}{(1-e^2)(m_{\rm p}+m_{\rm c})^{4/3}}.
\end{equation}

\noindent Geodetic precession leads to nonlinear rates of change in ($\lambda$, $\eta$). The time dependence of ($\lambda$, $\eta$) therefore requires that Equation \ref{eq:didtLT} -- and thus $\dot{x}_{\rm LT} = x\cot{i}(di/dt)_{\rm LT}^{\rm GR}$ -- also depend on time. This time dependence can be written as a set of coupled trigonometric equations \citep{dt92}:

\begin{align}
    \label{eq:cosl}
    \cos\lambda &= \cos\delta\cos i - \sin\delta\sin i\cos(\Omega_{\rm p}t + \Phi_0) \\
    \label{eq:cose}
    \cos\eta &= \sin\delta\sin(\Omega_{\rm p}t + \Phi_0) / \sin\lambda,
\end{align}

\noindent where $\delta = \cos^{-1}({\bf \hat{L}\cdot{\bf \hat{S}_{\rm p}}})$ is the misalignment angle between the senses of ${\bf L}$ and ${\bf S_{\rm p}}$ -- labeled with the ``hat" notation to indicate their respective unit vectors -- and $\Phi_0$ is a phase of orientation during the precession cycle.

The observed apsidal motion in pulsar-binary systems ($\dot{\omega} \equiv d\omega/dt$, where $\omega$ is the argument of periastron) is also expected to consist of additive terms associated with different mechanisms. For example, in the presently considered case of an isolated binary system, $\dot{\omega}_{\rm obs} = \dot{\omega}_{\rm int} + \dot{\omega}_\mu + \dot{\omega}_{\rm LT}$, where $\dot{\omega}_{\rm int}$ is the ``intrinsic" term associated with relativistic point-particle dynamics \citep{ds88}, while $\dot{\omega}_\mu$ is an ``extrinsic" bias due to proper motion of the binary system \citep{kop96}. However, unlike $\dot{x}_{\rm LT}$, $\dot{\omega}_{\rm LT}$ is bilinear in $\hat{\bf L}\cdot\hat{\bf S}_{\rm p} = \cos\delta$ and $\hat{\bf K}\cdot\hat{\bf S}_{\rm p}=\cos\lambda$, where $\hat{\bf K}$ is the unit vector pointing from the observer to the center of mass of the pulsar-binary system. While $\delta$ is constant, $\cos\lambda$ consists of constant and time-variable terms as shown in Equation \ref{eq:cosl} \citep{ds88}. This circumstance therefore means that $\dot{\omega}_{\rm LT}$ consists of constant and time-variable components, i.e., $\dot{\omega}_{\rm LT} = \dot{\omega}_{\rm LT,0} + \dot{\omega}_{\rm LT,t}$, where

\begin{align}
    \label{eq:omdotLT0}
    \dot{\omega}_{\rm LT,0} &= -\frac{4\pi GI_{\rm p}\cos\delta}{P_{\rm s}c^2a^3(1-e^2)^{3/2}}\bigg(2+\frac{3}{2}\frac{m_{\rm c}}{m_{\rm p}}\bigg), \\
    \label{eq:omdotLTt}
    \dot{\omega}_{\rm LT,t} &= +\frac{2\pi GI_{\rm p}\cot{i}\sin\delta}{P_{\rm s}c^2a^3(1-e^2)^{3/2}}\bigg(2+\frac{3}{2}\frac{m_{\rm c}}{m_{\rm p}}\bigg)\cos(\Omega_{\rm p}t + \Phi_0).
\end{align}

Past works have focused on the detection of first-order variations of $x$ and $\omega$ in known DNS systems \footnote{Pulsar timing is only sensitive to radial displacements of the pulsar, which do not depend on the longitude of ascending node ($\Omega$) that is also shown in Figure \ref{fig:angles}. The rate of change in $\Omega$ due to LT precession is therefore not directly measurable and is subsequently ignored in further discussion.}. However, the time dependence in Equations \ref{eq:didtLT} and \ref{eq:omdotLTt} suggests that higher-order variations arise after sufficient LT precession of the orbit has occurred \citep[e.g.,][]{wk99}. The full effect of LT precession can therefore be captured by integrating Equations \ref{eq:didtLT}, \ref{eq:aber}, and \ref{eq:omdotLTt}, which leads to a redefinition of the orbital timing model developed by \citep{dd86}:

\begin{align}
    \label{eq:ECCLT}
    e(t) &= e_0 + \Delta e_{\epsilon_{\rm p}}(t) + \dot{e}(t-T_0) \\
    \label{eq:A1LT}
    x(t) &= x_0 + \Delta x_{\rm LT}(t) + \Delta x_{\epsilon_{\rm p}}(t) + \dot{x}(t-T_0) \\
    \label{eq:OMLT}
    \omega(t) &= \omega_0 + \Delta \omega_{\rm LT}(t) + \bigg(\frac{P_{\rm b}\dot{\omega}_{\rm c}}{2\pi}\bigg)u(t),
\end{align}

\noindent where the integrated expressions, represented as $\Delta$ terms, are presented in Appendix \ref{app:model}, $T_0$ is the epoch of periastron passage, $\dot{\omega}_c = \dot{\omega}_{\rm int} + \dot{\omega}_\mu + \dot{\omega}_{\rm LT,0}$, and $u(t)$ is the true anomaly.

\begin{figure*}
    \centering
    \includegraphics[scale=0.75]{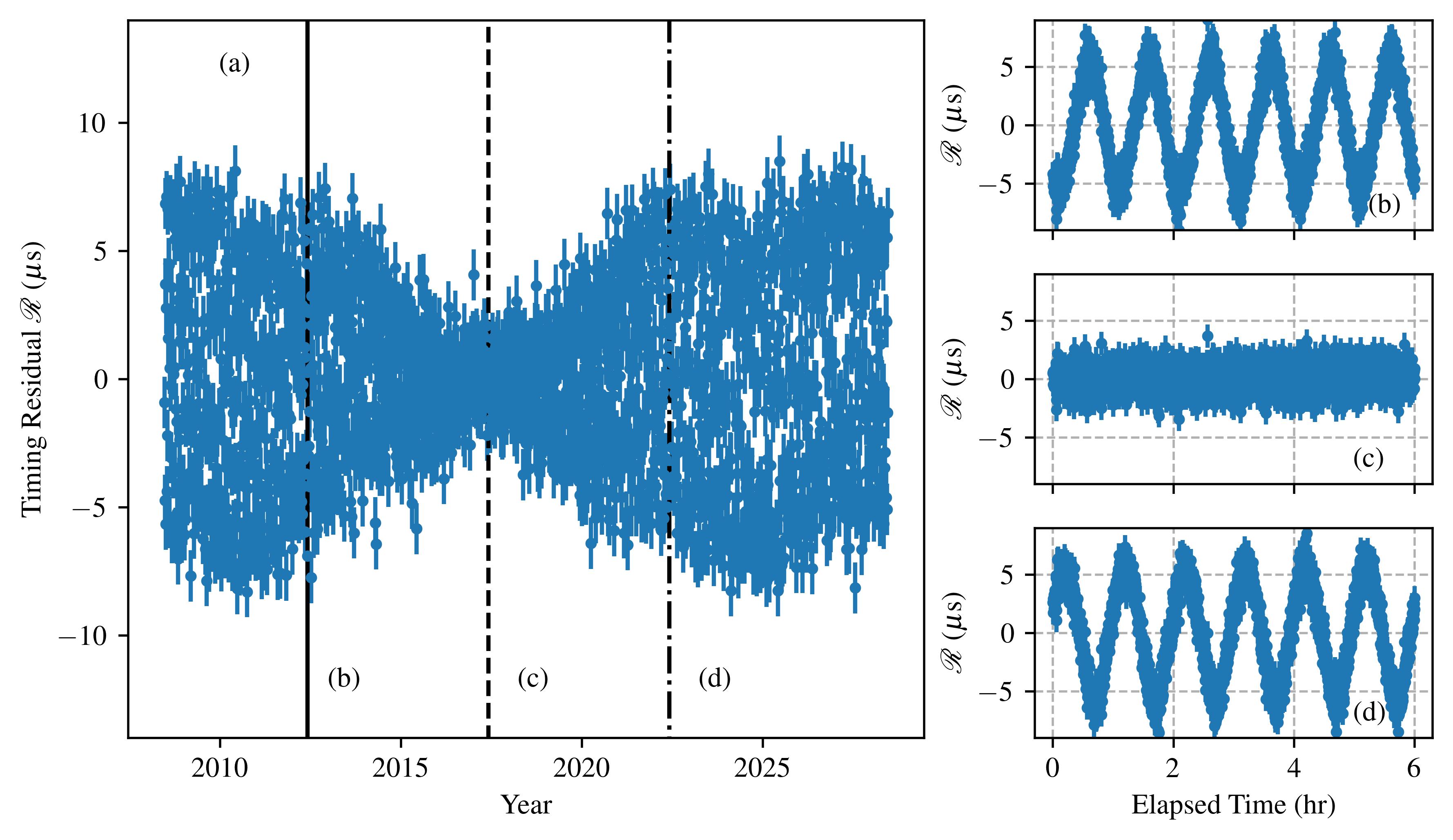}
    \caption{Simulated pulsar-timing residuals of a fictitious DNS binary pulsar with $P_{\rm b} = 1$ hr, described in \S\ref{sec:vars}, that exhibits LT precession but with $\dot{\omega}_{\rm c} = 0$ deg yr$^{-1}$ to illustrate only periodic LT variations. Panel (a) shows residuals over a 20-yr timespan. The panels (b,c,d) on the right show residuals evaluated over 6-hr time intervals at different times labeled in Panel (a), showing that timing effects from $\dot{\omega}_{\rm LT,t}$ vary significantly for this hypothetical binary pulsar.}
    \label{fig:resids}
\end{figure*}

According to Equations \ref{eq:ECCLT}--\ref{eq:OMLT}, DNS pulsars with $i < 90^\circ$ and $\delta > 0^\circ$ will exhibit \{$\Delta e_{\epsilon_{\rm p}}$, $\Delta x_{\rm LT}$, $\Delta \omega_{\rm LT}$\}, along with $\dot{\omega}_{\rm LT,0}$ that contributes to $\dot{\omega}_{\rm c}$. The aggregate effect from LT precession will therefore be a secular change in $\omega$ along with periodic modulations in \{$e$, $x$, $\omega$\}. The modulations from LT precession will have a timescale equal to $2\pi/\Omega_{\rm p}$. In order to visualize this overall effect, we used the open-source {\tt PINT} pulsar-timing software suite \citep{lsd+21} and extended its existing ``DDGR" binary timing model \citep{dd85,dd86} to include (a) second-PN-order corrections to all relativistic variations wherever relevant, and (b) Equations \ref{eq:didtLT}--\ref{eq:OMLT} into the parameterized expressions for \{$e$, $x$, $\omega$\}. The resultant model, which we hereafter refer to as ``DDGRI", contains ($I_{\rm p}$, $\delta$, $\Phi_0$) as degrees of freedom, along with the five Keplerian elements and ($m_{\rm c}$, $m_{\rm tot} = m_{\rm p}+m_{\rm c}$).

Using this DDGRI model, we simulated the timing proprerties of a fictitious DNS pulsar with $P_{\rm b} = 1$ hr, $a_{\rm p} = 0.81$ lt-s \footnote{The value of $a_{\rm p}$ was chosen to ensure that $\cot i \sim 0.1$ as required by the Keplerian mass function.}, $e=0.1$, $\delta=10^\circ$, and $m_{\rm c} = m_{\rm p} = 1.4\textrm{ M}_\odot$. Figure \ref{fig:resids} shows the difference (or ``residuals") in simulated pulsar times of arrival (TOAs) between a timing model where $I_{\rm p} = 1.25\times10^{45}\textrm{ gm cm}^2$ and another where $I_{\rm p} = 0\textrm{ gm cm}^2$, with the latter model corresponding to no LT precession; for both TOA simulations, we also assumed $\dot{\omega}_c = 0$ deg yr$^{-1}$ purely to illustrate effects arising from the nonlinear nature of LT precession. LT precession gradually arises as nonlinear orbital perturbations at the $O(1-10\textrm{ }\mu$s) level in TOA residuals for this fictitious DNS pulsar.

\begin{table*}[]
    \centering    
    \begin{tabular}{|l|r|r|r|r|r|r|r|r|r|r|l|}
       \hline\hline
       PSR  & $m_{\rm p}$ (M$_\odot$) & $m_{\rm c}$ (M$_\odot$) & $P_{\rm b}$ (hr) & $e$ & $i$ (deg) & $\delta$ (deg) & $\Phi_0$ (deg) & $\Omega_{\rm p}$ ($^\circ/$yr) & $\sigma_{\rm TOA}$ ($\mu$s)\footnote{For PSRs J0737$-$3039A and J1757$-$1854, the larger value corresponds to precision currently achieved with the GBT; the smaller value corresponds to the expected factor of $\sim$ 8 improvement in precision to be achieved with the DSA.} & $T$ (yr) & References \\
       \hline
       J0737$-$3039A & 1.38 & 1.24 & 2.45 & 0.08 & 89.3 & $<3.2$ & 45.0\footnote{This value is arbitrarily chosen as there is no observational constraint on $\Phi_0$.} & 4.7 & 6.0/0.75 & 26.7 & \cite{ksm+21,hkc+22} \\
       J1537+1155\footnote{This pulsar is historically referred to as PSR B1534+12.} & 1.33 & 1.34 & 10.0 & 0.27 & 77.7 & $27^{+3}_{-3}$ & $290^{+20}_{-20}$ & 0.5 & 2.0 & 39.2 &  \cite{sta04,fst10} \\
       J1757$-$1854 & 1.34 & 1.39 & 4.40 & 0.60 & 85.3 & $55.60^{+1.88}_{-2.01}$ & $169.40^{+1.80}_{-2.21}$ & 3.0 & 10.0/1.25 & 15.1 & \cite{cbc+23} \\
       J1915+1606\footnote{This pulsar is historically referred to as PSR B1913+16, or the ``Hulse-Taylor pulsar."} & 1.43 & 1.39 & 7.75 & 0.61 & 132.8 & $22^{+3}_{-8}$ & $55^{+5}_{-5}$ & 1.2 & 8.0 & 49.0 & \cite{kra98,wh16} \\ 
       J1946+2052 & 1.23 & 1.24 & 1.88 & 0.06 & 63.0 & $0.21^{+0.21}_{-0.10}$ & $138.18^{+17.39}_{-44.25}$ & 7.8 & 9.0 & 15.1 & \cite{mzk+24,mfs+25} \\
       \hline
    \end{tabular}
    \caption{A selection of orbital parameters for five compact DNS pulsars that have constraints or significant measurements of ($\delta$, $\Phi_0$) from confirmed geodetic precession. Reported statistical uncertainties reflect 68.3\% credibility intervals. Values with no listed uncertainties have reported precision at scales smaller than the significant digits listed in this table.}
    \label{tab:data}
\end{table*}

\section{Known DNS Pulsars}
\label{sec:knownpsrs}

There are $O(20)$ known DNS systems within the Milky Way Galaxy that contain at least one pulsar. Nearly half of these systems yield one or more orbital variations due to GR and can therefore be considered ``relativistic." However, only a small fraction of known DNS pulsars have constraints on $\delta$ obtained through analyses of pulse-shape and polarization variations \citep{dt92}. Moreover, a similarly small fraction of DNS systems exhibit timing precision at the $O(10\textrm{ }\mu\textrm{s})$ level or better. We therefore restricted an analysis of the known DNS pulsars to sources that have (a) timing baselines exceeding at least 10 yr, (b) TOA uncertainties less than or equal to $10\textrm{ }\mu$s, and (c) reported constraints \{$\delta$, $\Phi_0$\}. The application of these criteria yields five sources: PSR J0737$-$3039A; J1537+1155; J1757$-$1854; J1915+1606; and J1946+2052. A summary of relevant measurements for these five DNS systems is provided in Table \ref{tab:data}.

For each known pulsar, we conducted {\tt PINT} simulations using the DDGRI model and in a manner similar to that first described at the end of \S\ref{sec:vars}. However, for each pulsar, we simulated a number of TOA data sets of increasing $T$, up to the year 2040, to determine if the nonlinear effects from LT precession are detectable in the near future. We implemented the following, additional simulation steps to ensure robustness in constraining the presence of LT precession in pulsar-timing data:

\begin{itemize}
    \item each DDGRI simulation calculated all expected orbital perturbations from GR as encoded in Equations \ref{eq:ECCLT}--\ref{eq:OMLT}, including the full $\dot{\omega}_c$ term, using the parameters listed in Table \ref{tab:data};
    \item each simulation assumed the same starting epoch that matches the date of first observation reported for each pulsar in Table \ref{tab:data};
    \item each simulated observation was assumed to last 3 hours and itself consisted of a number of TOAs within this interval that each spanned 5\% of the pulsar's orbit;
    \item each TOA possessed an uncertainty equal to the value of $\sigma_{\rm RMS}$ in Table \ref{tab:data};
    \item each simulation consisted of a number of observations with a cadence of 30 days.
\end{itemize}

\noindent For J0737$-$3039 and J1757$-$1854, the simulations instead assumed a cadence of 15 days and accounted for expected improvement in timing precision with the forthcoming operation of the Deep Synoptic Array \citep[DSA;][]{hrw+19}, assuming that it begins observations in January 2028. While stringent, the heightened cadence matches that achieved in practice for J0737$-$3039A \citep{ksm+21}.

\begin{figure*}
    \centering
    \includegraphics[width=0.49\textwidth]{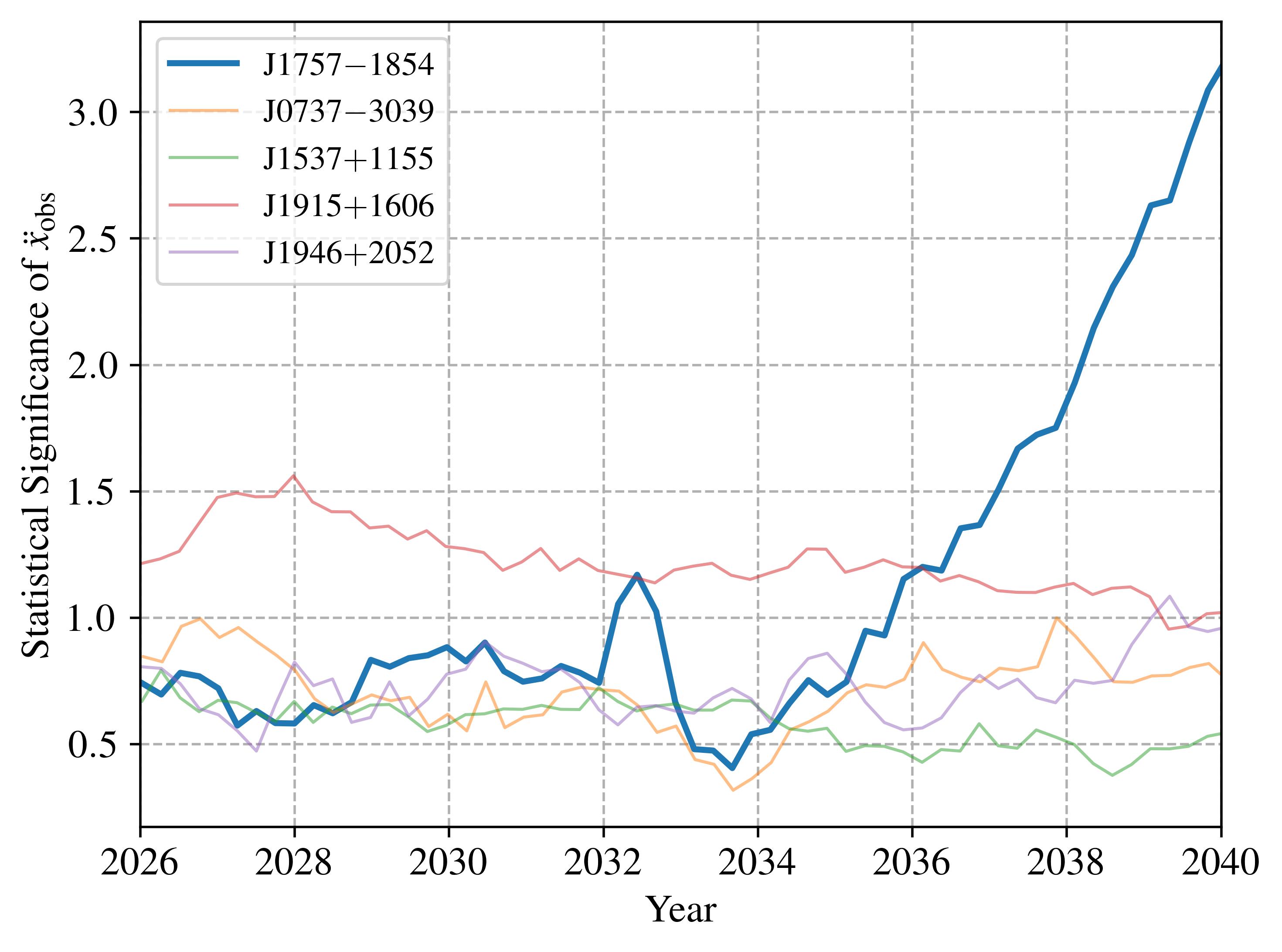}
    \includegraphics[width=0.49\textwidth]{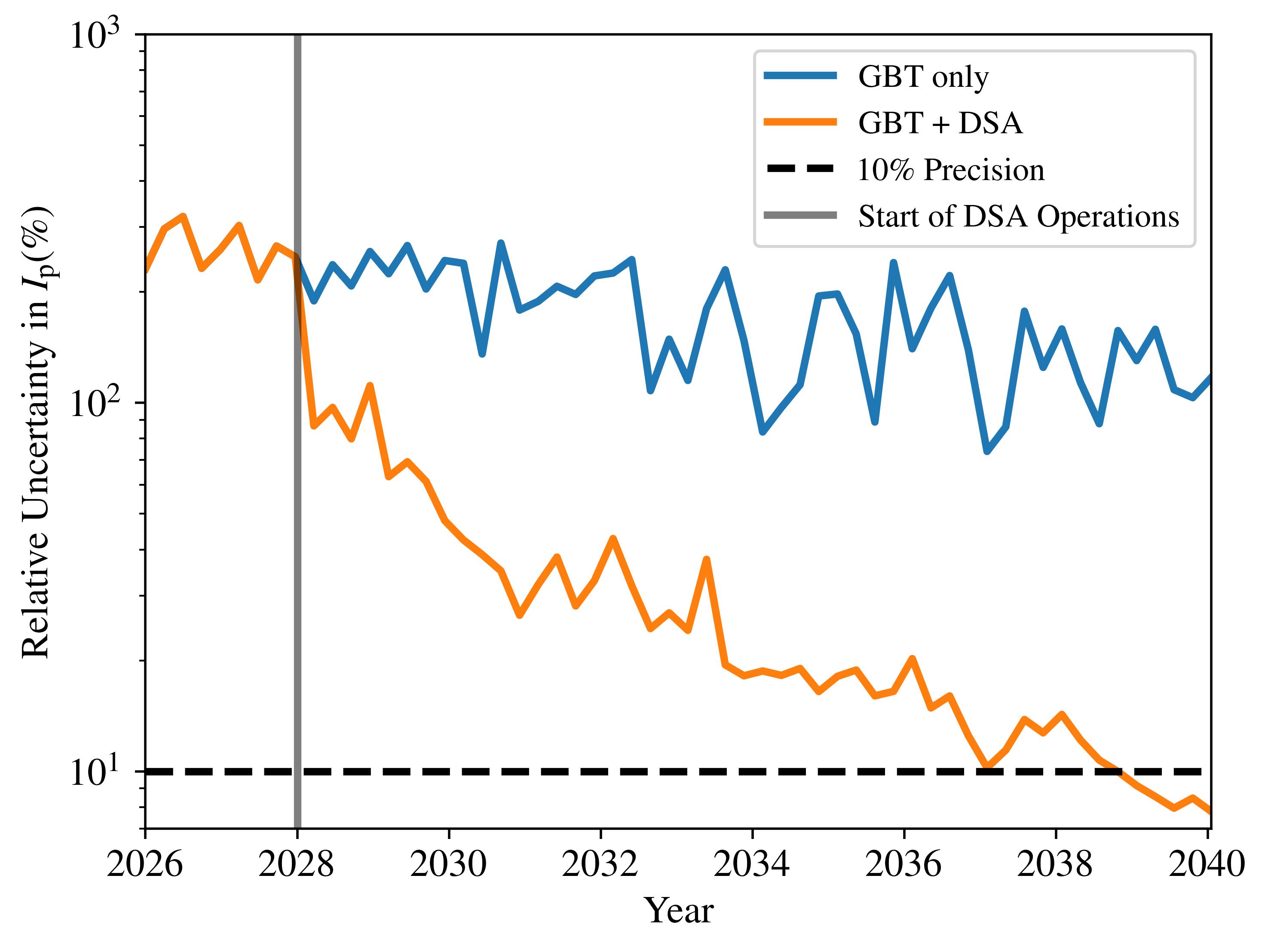}
    \caption{Projected impacts of LT precession in the timing properties of the known DNS pulsars listed in Table \ref{tab:data}. ({\it Left.}) Simulated measurements of $\ddot{x}_{\rm obs, LT} = \ddot{x}_{\rm obs} - \ddot{x}_{\epsilon_{\rm p}}$ from {\tt PINT} when modeling the simulated TOAs discussed in Section \ref{sec:knownpsrs}. Only PSR J1757$-$1854 exhibits a statistically significant measurement of $\ddot{x}_{\rm obs, LT}$ that is also consistent with the expected value arising from LT precession. ({\it Right.}) Relative uncertainty in $I_{\rm p}$ in PSR J1757$-$1854, obtained from simulations that do and do not account for the improvement in timing decision from eventual DSA observations assumed to begin in 2028. These projections use the \{$\delta$, $\Phi_0$\} measurements and uncertainties from \citep{cbc+23} as Bayesian prior information.}
    \label{fig:results}
\end{figure*}

To detect the presence of nonlinearity from LT precession, we  analyzed all simulated TOA data sets using a modified version of the phenomenological ``DD" binary-timing model. The original DD model accounts for PN effects only as first-order variations which, along with the five Keplerian elements, introduces \{$m_{\rm c}$, $\sin{i}$,  $\dot{\omega}_{\rm obs}$, $\dot{P}_{\rm b, obs}$, $\dot{e}_{\rm obs}$, $\dot{x}_{\rm obs}$\} as free parameters. However, we modified the DD model to replace the $\Delta$ terms in Equations \ref{eq:ECCLT}--\ref{eq:OMLT} with second-order Taylor expansions in time. This extension of the DD model introduces the second time-derivatives \{$\ddot{e}$, $\ddot{x}$, $\ddot{\omega}$\} as additional degrees of freedom; a significant measurement of any second time-derivative would occur only if sufficient LT precession induces nonlinear variations in $e$, $x$, and/or $\omega$.

We fitted this extended DD model against all simulated TOAs to produce a set of best-fit Keplerian elements and measurements of all PN variations, including \{$\ddot{e}_{\rm obs}$, $\ddot{x}_{\rm obs}$, $\ddot{\omega}_{\rm obs}$\}. The left-hand panel of Figure \ref{fig:results} shows the statistical significance of $\ddot{x}_{\rm obs}$ over time for all five pulsars listed in Table \ref{tab:data} \footnote{Each curve in these figures represents the average of 10 simulations that each generate different realizations of statistically white noise with standard deviation equal to $\sigma_{\rm RMS}$. This averaging is performed to minimize short-term fluctuations due to noise.}. The measurement of $\ddot{x}_{\rm obs}$ -- as well as \{$\ddot{e}_{\rm obs}$, $\ddot{\omega}_{\rm obs}$\} -- is not significant for PSRs J0737$-$3039A, J1537$+$1155, J1915$+$1606, and J1946$+$2052. These findings are consistent with expectations as both J0737$-$3039A and J1946$+$2052 have $\delta \sim 0^\circ$, and $T < 2\pi/\Omega_{\rm p}$ for J1537$+$1155 and J1915$+$1606; these circumstances indicate that LT precession will manifest as one or more linear variations and thus contribute only to \{$\dot{e}_{\rm obs}$, $\dot{x}_{\rm obs}$, $\dot{\omega}_{\rm obs}$\}.

However, the left-hand panel of Figure \ref{fig:results} also shows that PSR J1757$-$1854 will exhibit $\ddot{x}_{\rm obs}$ with growing significance over time. The mean, best-fit value of $\ddot{x}_{\rm obs} = -1.5(5)\times10^{-23}\textrm{ lt-s s}^{-2}$ by 2040, which agrees with the expected value of $\ddot{x}^{\rm GR}_{\rm LT} \approx -2.1\times10^{-23}\textrm{ lt-s s}^{-2}$. We interpret this agreement as a simulated detection of nonlinear perturbations from LT precession in PSR J1757$-$1854 by this time. This lone detection of $\ddot{x}_{\rm LT}$ reflects the uniqueness of PSR J1757$-$1854, as it possesses the shortest $T$ and largest $\sigma_{\rm RMS}$ but is nonetheless considered to be among the most ``relativistic" DNS pulsars due to its low $P_{\rm b}$ and large $e$. Moreover, the improved timing precision from DSA observations likely strengthens the ability to resolve LT precession. PSR J1757$-$1854 therefore appears to be the best candidate for detecting nonlinear variations from LT precession in the near future.

In the absence of external information, a detection of this nonlinearity will yield weak constraints on ($I_{\rm p}$, $\delta$, $\Phi_0$) due to the low precision with which $\ddot{x}$ is expected to be measured by 2040. Conversely, we can instead obtain constraints on $I_{\rm p}$ if existing measurements of ($\delta$, $\Phi_0$) are used as prior information; the statistical significance of $I_{\rm p}$ will therefore reflect the overall significance of LT precession in TOA data. To explore this case, we conducted an additional set of simulations for J1757$-$1854 to explore the achievable significance of $I_{\rm p}$. We used the Markov Chain Monte Carlo (MCMC) capabilities within {\tt PINT} for Bayesian probabilistic sampling of DDGRI parameters against the simulated, noise-injected TOA data sets described earlier in this section. We also used a version of the DDGRI model that does not model $\dot{\omega}_{\rm LT,0}$ and instead allows for the estimate of $\dot{\omega}_{\rm c}$; this parameterization is best for assessing the prominence of timing signatures associated with the periodic LT variations. Moreover, we assumed Gaussian priors for ($\delta$, $\Phi_0$) using the values reported in Table \ref{tab:data}, and employed a log-uniform prior in $I_{\rm p}$ over six orders of magnitude to ensure conservative estimation of posterior statistics. All other parameters -- the five Keplerian orbital elements and \{$m_{\rm c}$, $m_{\rm tot}$, $\dot{\omega}_{\rm c}$\} -- were sampled assuming Gaussian priors centered on their simulation values and with widths two times greater than least-squares estimates of their statistical uncertainties.

The results of our MCMC simulations for PSR J1757$-$1854 are shown in the right-hand panel of Figure \ref{fig:results}. The constraints on $I_{\rm p}$ clearly improve over the next decade, and especially once TOA measurements begin with the DSA. The relative uncertainty on $I_{\rm p}$ reaches 10\% by 2040 when using the DDGRI model, even when using a broad prior on $I_{\rm p}$ and simultaneously sampling the ($\delta$, $\Phi_0$) parameters using priors based on existing measurements. For comparison, we show constraints on $I_{\rm p}$ obtained using a similar MCMC experiment but neglecting the impact of the DSA, i.e., only collecting TOAs with an instrument with sensitivity equal to the Green Bank Telescope (GBT). The constraint on $I_{\rm p}$ differs by one order of magnitude between a GBT or DSA study of J1757$-$1854.

\section{Discussion}
\label{sec:discussion}

Nonlinear LT variations in known DNS pulsars will require extended timing baselines to separately measure \{$I_{\rm p}$, $\delta$, $\Phi_0$\} from the pulsar-timing signatures alone. Independent estimates of ($\delta$, $\Phi_0$) through studies of ``pulse structure" -- long-term changes in pulse shape and polarization \citep[e.g.,][]{dt92} -- are therefore crucial for using the timing-based detection of LT precession to constrain $I_{\rm p}$ in these known systems, as was shown for PSR J1757$-$1854 in \S\ref{sec:knownpsrs}. Nevertheless, detections of nonlinear orbital variations from LT precession can serve other purposes when assuming a neutron-star equation of state and therefore a value of $I_{\rm p}$ \citep[e.g.,][]{bs05}. 

For example, a detection of LT precession through pulsar timing can be used to check for self-consistency in ($\delta$, $\Phi_0$) inferred from pulse-structure analysis. Furthermore, the DDGRI model can be turned into a phenomenological form where \{$I_{\rm p}$, $\delta$\} are replaced with parameters that quantify the unique amplitudes of all LT variations and the angular frequency $(\dot{\Phi}_{\rm LT})$ of the periodic terms. This generalization would provide three new testable parameters of GR for systems with $\delta > 0^\circ$:

\begin{align}
    \dot{\omega}_{\rm LT,0} &= \lambda_{\rm LT}\dot{\omega}^{\rm GR}_{\rm LT,0}, \\
    \Delta \omega_{\rm LT} &= \mu_{\rm LT} \Delta \omega_{\rm LT}^{\rm GR}, \\
    \Delta x_{\rm LT} &= \mu_{\rm LT} \Delta x_{\rm LT}^{\rm GR},
\end{align}

\noindent where $\lambda_{\rm LT} = \mu_{\rm LT} = 1$ and $\dot{\Phi}_{\rm LT} = \Omega_{\rm p}$ in GR. For systems with $\delta \approx 0^\circ$, the detectable variations from LT precession reduce to $\dot{\omega}_{\rm LT,0}$ alone and therefore yield a single, amplitude-only test of LT precession. This test is currently possible for PSR J0737$-$3039A, where $\lambda_{\rm LT} = 0.7 \pm 0.9 $ \citep[68\% confidence;][]{ksm+21}.

The analyses presented in \S\ref{sec:knownpsrs} do not simulate sources of statistically ``red" noise in pulsar TOAs. Typical red noise in TOAs include stochastic variations in pulse dispersion \citep[e.g.,][]{mml+17}, and other sources of ``timing noise" related to the instability in pulsar rotation \citep[e.g,][]{abh23}. The presence of red noise limits the ability to test GR using DNS pulsars if left untreated \citep[e.g.,][]{hu25}. However, a variety of methods exist to simultaneously model red noise through phenomenological means \citep[e.g.,][]{rzs+23}. We explored the impact of modeling pulse-dispersion variations when simultaneously estimating $I_{\rm p}$ for PSR J1757$-$1854 in Appendix \ref{app:noise} and found that the required modeling only reduces the statistical significance in $I_{\rm p}$ to $O(20\%)$ by 2040. This result indicates that statistically-red dispersion variations will likely not prevent a significant measurement of $I_{\rm p}$ when using our model; further exploration of red noise and its impact in measuring $I_{\rm p}$ will be the subject of future work.

\begin{acknowledgements}
    E.~F. is grateful to Norbert Wex, Carolyn Raithel,  Paulo C.~C. Freire, and Vivek. V. Krishan for valuable discussion. E.~F. is supported by the National Science Foundation under grant AST-2407399.
\end{acknowledgements}

\appendix

\section{Implementation of LT Precession}
\label{app:model}

As discussed in the main text, timing perturbations due to LT precession arise as four distinct terms. One of these terms is the first-order variation shown in Equation \ref{eq:omdotLT0} that contributes to the constant apsidal motion of the binary system. The second term is Equation \ref{eq:omdotLTt}, which demonstrates a periodic variation in $\omega$ over time due to LT precession. Direct integration of Equation \ref{eq:omdotLTt} yields

\begin{align}
    \Delta \omega_{\rm LT} &= \int_0^t \dot\omega_{\rm LT}(t')dt' \\
    &\approx \frac{2\pi GI_{\rm p}\cot{i}\sin\delta}{\Omega_{\rm p}P_{\rm s}c^2a^3(1-e^2)^{3/2}}\bigg(2+\frac{3}{2}\frac{m_{\rm c}}{m_{\rm p}}\bigg) \\ \nonumber
    & \phantom{{} = AAA} \times \big[\sin(\Omega_{\rm p } t + \Phi_0) - \sin\Phi_0\big],
\end{align}

\noindent where the approximation sign reflects the assumption that the linear PN variations in the orbital elements are negligible in comparison to the short-term evolution from LT precession. The third term is Equation \ref{eq:didtLT}, which can be put as an explicit function of time by noting from Equation \ref{eq:cose} that $\sin\lambda\cos\eta = \sin\delta\sin(\Omega_{\rm p}t + \Phi_0).$ Direct integration of Equation \ref{eq:didtLT} similarly yields

\begin{align}
    \Delta x_{\rm LT} &\approx x\cot i  \int_0^t \dot{i}_{\rm LT}(t')dt' \\
    &\approx -\frac{2\pi GI_{\rm p}x\cot{i}\sin\delta}{\Omega_{\rm p}P_{\rm s}c^2a^3(1-e^2)^{3/2}}\bigg(2+\frac{3}{2}\frac{m_{\rm c}}{m_{\rm p}}\bigg) \\ \nonumber
    & \phantom{{} = AAA} \times \big[\cos(\Omega_{\rm p } t + \Phi_0) - \cos\Phi_0\big],
\end{align}

The perturbations in \{$e$, $x$\} due to changing aberration of the pulsed radio signal -- represented by Equations \ref{eq:xdot} and \ref{eq:aber}  -- do not arise from LT precession. However, Equation \ref{eq:aber} is a function of $(\lambda, \eta)$ which vary due to LT precession, and is thus itself a function of time. The trigonometric combinations in Equation \ref{eq:aber} can be put into explicit functions of time using Equations \ref{eq:cosl}--\ref{eq:cose}. Direct integration of Equation \ref{eq:aber} yields the following closed-form expression,

\begin{align}
    \Delta {\epsilon_{\rm p}} &= -\frac{P_{\rm s}}{P_{\rm b}\sqrt{1-e^2}}\times\bigg[ \\ \nonumber
    & \phantom{{} = AAAA+}A\bigg(\frac{y}{2[1-y^2]}+\frac{1}{4}\ln\frac{1+y}{1-y}\bigg) \\ \nonumber
    & \phantom{{} = AAA} + B\bigg(C\bigg\{\frac{y}{2[1-y^2]}+\frac{1}{4}\ln\frac{1+y}{1-y}\bigg\} \\ \nonumber 
    & \phantom{{} = AAA} - \bigg\{\frac{1}{2[1-y^2]}\bigg\}\bigg) \\ \nonumber 
    & \phantom{{} = AAA} + D\bigg\{C^2\bigg(\frac{y}{2[1-y^2]}+\frac{1}{4}\ln\frac{1+y}{1-y}\bigg) \\ \nonumber
    & \phantom{{} = AAA} - 2C \bigg(\frac{1}{2[1-y^2]}\bigg) \\ \nonumber 
    & \phantom{{} = AAA} + \bigg(\frac{y}{2[1-y^2]}-\frac{1}{4}\ln\frac{1+y}{1-y}\bigg)\bigg\} \\ \nonumber
    & \phantom{{} = AAA} - \frac{1}{4}D\ln\frac{1+y}{1-y}\bigg]\bigg|_{y(t=0)}^{y(t)},
\end{align}

\noindent where the following coefficients are related to trigonometric terms:

\begin{align}
    A &= 2\cos{i}\cos^2\delta, \\ \nonumber
    B &= 2\cos\delta(\cot^2{i}-1), \\ \nonumber 
    C &= \cos\delta\cos{i}, \\ \nonumber
    D &= -2\cos{i}/\sin^2{i}, \\ \nonumber
    y &= \cos\lambda.
\end{align}

\noindent The perturbations in \{$e$, $x$\} due to aberration can therefore be written as $\Delta e_{\epsilon_p} \approx e\Delta {\epsilon_{\rm p}}$ and $\Delta x_{\epsilon_p} \approx x\Delta {\epsilon_{\rm p}}$, respectively.

\section{The Impact of Red Noise in Measuring LT Precession}
\label{app:noise}

The simulations in the main text relied on two key properties: an accurate measure of $\sigma_{\rm TOA}$ for each pulsar; and the assumption that TOA residuals exhibit random fluctuations consistent with white noise. In reality, these properties may depend on time and/or electromagnetic frequency ($\nu$) due to several factors that are specific to each pulsar. For example, the spectral energy distribution of pulsars is known to vary throughout the radio spectrum and often in power-law form \citep[e.g.,][]{blv+13}. For the purposes of our study, this fact will require $\sigma_{\rm TOA} \equiv \sigma_{\rm TOA}(\nu)$. We circumvented this dependency in simulations presented in the main text by assuming observed pulsations were averaged across the entire range of sampled radiation. The values of $\sigma_{\rm TOA}$ in Table \ref{tab:data} were estimated from published data by averaging TOA uncertainties evaluated over a number of electromagnetic-frequency channels ($N_\nu$).

These single-TOA simulations also allowed us to ignore other  ``chromatic" (i.e., frequency-dependent) effects, such as cold-plasma pulse dispersion that is ubiquitous in pulsar observations. The dispersion time delay $\Delta t_{\rm DM} \propto {\rm DM}$, where DM $= \int_0^d n_e(l)dl$ is the ``dispersion measure", $d$ is the pulsar distance, and $n_{e}$ is the electron number density along the line of sight $l$. The DM can be measured using TOAs for a single pulse resolved over a frequency bandwidth with $N_\nu>1$.  Pulse dispersion is known to undergo stochastic variations over time as $l$ (and therefore $n_e$) changes due to relative motion \citep[e.g.,][]{mml+17}. Therefore, robust simulations must account for the fact that ${\rm DM} \equiv {\rm DM}(t)$ and its temporal fluctuations are consistent with the statistics of red noise.

We performed a separate set of simulations on PSR J1757$-$1854 to explore the impact of red noise, specifically manifested as red-like DM variations, in measuring $I_{\rm p}$. We used an observing strategy similar to that used by \citep{cbc+23}, where $N_\nu = 4$ and TOAs are evaluated within the 1700-2300 MHz range of the radio spectrum. While \citep{cbc+23} measured ${\rm DM} = 378.2145(6)$ pc cm$^{-3}$, they did not measure any significant DM variations in their 6-yr data set; this constancy suggests that the first-order variation in DM can be limited to be $\dot{\rm DM} \le 10^{-4}\textrm{ pc cm}^{-3}\textrm{ yr}^{-1}$. We nonetheless assumed a polynomial variation in DM described by a Taylor expansion for the purpose of testing its impact on our inference of $I_{\rm p}$. We arbitrarily set $\dot{\rm DM} = 10^{-3}$ pc cm$^{-3}\textrm{ yr}^{-1}$, $\ddot{\rm DM} = 2\times10^{-4}$ pc cm$^{-3}\textrm{ yr}^{-2}$, and $\dddot{\rm DM} = 3\times10^{-5}$ pc cm$^{-3}\textrm{ yr}^{-3}$.

\begin{figure}
    \centering
    \includegraphics[width=\linewidth]{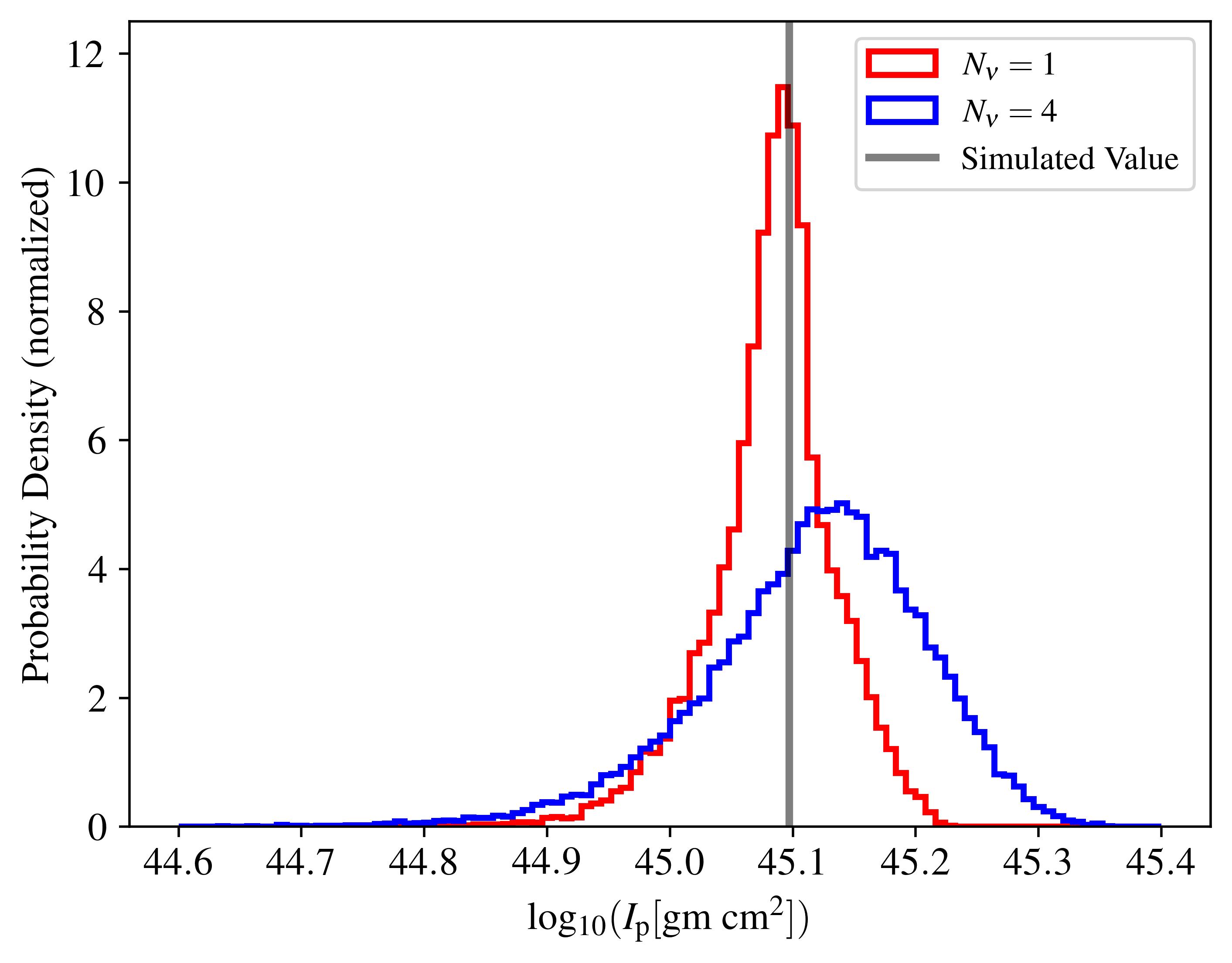}
    \caption{Realization-averaged posterior distributions of $I_{\rm p}$ for MCMC simulations that assume a constant DM ($N_\nu=1$) and directly model polynomial variations in DM ($N_\nu=4$).}
    \label{fig:I1_DMvars}
\end{figure}

We then performed a modified MCMC simulation to constrain $I_{\rm p}$ similar to that described in the main text, but including \{DM, $\dot{\rm DM}$, $\ddot{\rm DM}$, $\dddot{\rm DM}$\} as degrees of freedom with Gaussian prior distributions. The posterior distribution of $I_{\rm p}$ for simulations with $N_\nu=4$, for data sets spanning up to 2040, is shown in Figure \ref{fig:I1_DMvars}; also shown in this figure is the same distribution obtained for the same timespan but when $N_\nu=1$. As done in the main text, the posterior distributions are obtained by performing 10 separate MCMC simulations, each with a different realization of white noise in TOA residuals, and averaging the resultant distributions to minimize fluctuations from injected noise. These results indicate that $I_{\rm p}$ will be measured at the 15--20\% level when $N_\nu=4$ by 2040, which is a factor of $\sim$ 2 larger than the $N_\nu=1$ case presented in the main text.

The results in this Appendix indicate that stochastic DM variations, when directly modeled, will not prevent a statistically significant measurement of $I_{\rm p}$. However, the inclusion of red-like noise processes does partially lower the significance relative to the constant-DM case presented in the main text. Therefore, additional studies on the role that red noise will play in measuring $I_{\rm p}$ are warranted and will be the subject of future works.

An atypical source of red noise, unique to pulsars in highly relativistic binary systems, is the long-term change in the pulse profile due to changing orientation induced by geodetic precession. This secular change is expected to produce stochastic variations in $P_{\rm s}$ \citep{kws03}. Nearly all known DNS pulsars, including those presented in Table \ref{tab:data}, exhibit slowly varying profiles whose resultant timing variations can be modeled with a polynomial expansion of $P_{\rm s}$. By contrast, the relativistic PSR J1906+0746 shows extreme timing noise due to substantial pulse-profile evolution, with a magnitude well in excess of $O(10\textrm{ }\mu$s) \citep{vks+15}. This variation is likely due to the unusually high value of $\delta = 104(9)^\circ$ that leads to substantial precession of the pulsar's light cone; the observed variations and large $\delta$ are so extreme that the pulsar's light cone is expected to precess out of our line of sight as early as 2028 \cite{dkl+19}. However, the status of this pulsar as residing in a DNS system remains ambiguous and leaves the possibility that variations arise from classical spin-orbit coupling due to quadrupole-order interactions between the pulsar and white-dwarf companion \cite{vsk+26}.

\bibliography{journals,apssamp}

\begin{thebibliography}{39}%
\makeatletter
\providecommand \@ifxundefined [1]{%
 \@ifx{#1\undefined}
}%
\providecommand \@ifnum [1]{%
 \ifnum #1\expandafter \@firstoftwo
 \else \expandafter \@secondoftwo
 \fi
}%
\providecommand \@ifx [1]{%
 \ifx #1\expandafter \@firstoftwo
 \else \expandafter \@secondoftwo
 \fi
}%
\providecommand \natexlab [1]{#1}%
\providecommand \enquote  [1]{``#1''}%
\providecommand \bibnamefont  [1]{#1}%
\providecommand \bibfnamefont [1]{#1}%
\providecommand \citenamefont [1]{#1}%
\providecommand \href@noop [0]{\@secondoftwo}%
\providecommand \href [0]{\begingroup \@sanitize@url \@href}%
\providecommand \@href[1]{\@@startlink{#1}\@@href}%
\providecommand \@@href[1]{\endgroup#1\@@endlink}%
\providecommand \@sanitize@url [0]{\catcode `\\12\catcode `\$12\catcode
  `\&12\catcode `\#12\catcode `\^12\catcode `\_12\catcode `\%12\relax}%
\providecommand \@@startlink[1]{}%
\providecommand \@@endlink[0]{}%
\providecommand \url  [0]{\begingroup\@sanitize@url \@url }%
\providecommand \@url [1]{\endgroup\@href {#1}{\urlprefix }}%
\providecommand \urlprefix  [0]{URL }%
\providecommand \Eprint [0]{\href }%
\providecommand \doibase [0]{https://doi.org/}%
\providecommand \selectlanguage [0]{\@gobble}%
\providecommand \bibinfo  [0]{\@secondoftwo}%
\providecommand \bibfield  [0]{\@secondoftwo}%
\providecommand \translation [1]{[#1]}%
\providecommand \BibitemOpen [0]{}%
\providecommand \bibitemStop [0]{}%
\providecommand \bibitemNoStop [0]{.\EOS\space}%
\providecommand \EOS [0]{\spacefactor3000\relax}%
\providecommand \BibitemShut  [1]{\csname bibitem#1\endcsname}%
\let\auto@bib@innerbib\@empty
\bibitem [{\citenamefont {{Bailes}}(1988)}]{bai88}%
  \BibitemOpen
  \bibfield  {author} {\bibinfo {author} {\bibfnamefont {M.}~\bibnamefont
  {{Bailes}}},\ }\bibfield  {title} {\bibinfo {title} {{Geodetic precession in
  binary pulsars.}},\ }\href@noop {} {\bibfield  {journal} {\bibinfo  {journal}
  {Astro. \& Astrop.}\ }\textbf {\bibinfo {volume} {202}},\ \bibinfo {pages}
  {109} (\bibinfo {year} {1988})}\BibitemShut {NoStop}%
\bibitem [{\citenamefont {{Damour}}\ and\ \citenamefont
  {{Schafer}}(1988)}]{ds88}%
  \BibitemOpen
  \bibfield  {author} {\bibinfo {author} {\bibfnamefont {T.}~\bibnamefont
  {{Damour}}}\ and\ \bibinfo {author} {\bibfnamefont {G.}~\bibnamefont
  {{Schafer}}},\ }\bibfield  {title} {\bibinfo {title} {{Higher-order
  relativistic periastron advances and binary pulsars.}},\ }\href
  {https://doi.org/10.1007/BF02828697} {\bibfield  {journal} {\bibinfo
  {journal} {Nuo. Cim. B Ser.}\ }\textbf {\bibinfo {volume} {101B}},\ \bibinfo
  {pages} {127} (\bibinfo {year} {1988})}\BibitemShut {NoStop}%
\bibitem [{\citenamefont {{Lyne}}\ \emph {et~al.}(2004)\citenamefont {{Lyne}},
  \citenamefont {{Burgay}}, \citenamefont {{Kramer}}, \citenamefont
  {{Possenti}}, \citenamefont {{Manchester}}, \citenamefont {{Camilo}},
  \citenamefont {{McLaughlin}}, \citenamefont {{Lorimer}}, \citenamefont
  {{D'Amico}}, \citenamefont {{Joshi}}, \citenamefont {{Reynolds}},\ and\
  \citenamefont {{Freire}}}]{lbk+04}%
  \BibitemOpen
  \bibfield  {author} {\bibinfo {author} {\bibfnamefont {A.~G.}\ \bibnamefont
  {{Lyne}}}, \bibinfo {author} {\bibfnamefont {M.}~\bibnamefont {{Burgay}}},
  \bibinfo {author} {\bibfnamefont {M.}~\bibnamefont {{Kramer}}}, \bibinfo
  {author} {\bibfnamefont {A.}~\bibnamefont {{Possenti}}}, \bibinfo {author}
  {\bibfnamefont {R.~N.}\ \bibnamefont {{Manchester}}}, \bibinfo {author}
  {\bibfnamefont {F.}~\bibnamefont {{Camilo}}}, \bibinfo {author}
  {\bibfnamefont {M.~A.}\ \bibnamefont {{McLaughlin}}}, \bibinfo {author}
  {\bibfnamefont {D.~R.}\ \bibnamefont {{Lorimer}}}, \bibinfo {author}
  {\bibfnamefont {N.}~\bibnamefont {{D'Amico}}}, \bibinfo {author}
  {\bibfnamefont {B.~C.}\ \bibnamefont {{Joshi}}}, \bibinfo {author}
  {\bibfnamefont {J.}~\bibnamefont {{Reynolds}}},\ and\ \bibinfo {author}
  {\bibfnamefont {P.~C.~C.}\ \bibnamefont {{Freire}}},\ }\bibfield  {title}
  {\bibinfo {title} {{A Double-Pulsar System: A Rare Laboratory for
  Relativistic Gravity and Plasma Physics}},\ }\href
  {https://doi.org/10.1126/science.1094645} {\bibfield  {journal} {\bibinfo
  {journal} {Science}\ }\textbf {\bibinfo {volume} {303}},\ \bibinfo {pages}
  {1153} (\bibinfo {year} {2004})}\BibitemShut {NoStop}%
\bibitem [{\citenamefont {{Lattimer}}\ and\ \citenamefont
  {{Schutz}}(2005)}]{bs05}%
  \BibitemOpen
  \bibfield  {author} {\bibinfo {author} {\bibfnamefont {J.~M.}\ \bibnamefont
  {{Lattimer}}}\ and\ \bibinfo {author} {\bibfnamefont {B.~F.}\ \bibnamefont
  {{Schutz}}},\ }\bibfield  {title} {\bibinfo {title} {{Constraining the
  Equation of State with Moment of Inertia Measurements}},\ }\href
  {https://doi.org/10.1086/431543} {\bibfield  {journal} {\bibinfo  {journal}
  {ApJ}\ }\textbf {\bibinfo {volume} {629}},\ \bibinfo {pages} {979} (\bibinfo
  {year} {2005})}\BibitemShut {NoStop}%
\bibitem [{\citenamefont {{Hu}}\ and\ \citenamefont {{Freire}}(2024)}]{hf24}%
  \BibitemOpen
  \bibfield  {author} {\bibinfo {author} {\bibfnamefont {H.}~\bibnamefont
  {{Hu}}}\ and\ \bibinfo {author} {\bibfnamefont {P.~C.~C.}\ \bibnamefont
  {{Freire}}},\ }\bibfield  {title} {\bibinfo {title} {{Measuring the
  Lense{\textendash}Thirring Orbital Precession and the Neutron Star Moment of
  Inertia with Pulsars}},\ }\href {https://doi.org/10.3390/universe10040160}
  {\bibfield  {journal} {\bibinfo  {journal} {Universe}\ }\textbf {\bibinfo
  {volume} {10}},\ \bibinfo {eid} {160} (\bibinfo {year} {2024})}\BibitemShut
  {NoStop}%
\bibitem [{\citenamefont {{Lattimer}}\ and\ \citenamefont
  {{Prakash}}(2016)}]{lp16}%
  \BibitemOpen
  \bibfield  {author} {\bibinfo {author} {\bibfnamefont {J.~M.}\ \bibnamefont
  {{Lattimer}}}\ and\ \bibinfo {author} {\bibfnamefont {M.}~\bibnamefont
  {{Prakash}}},\ }\bibfield  {title} {\bibinfo {title} {{The equation of state
  of hot, dense matter and neutron stars}},\ }\href
  {https://doi.org/10.1016/j.physrep.2015.12.005} {\bibfield  {journal}
  {\bibinfo  {journal} {Phys. Rep.}\ }\textbf {\bibinfo {volume} {621}},\
  \bibinfo {pages} {127} (\bibinfo {year} {2016})}\BibitemShut {NoStop}%
\bibitem [{\citenamefont {{McMillan}}(2017)}]{mcm17}%
  \BibitemOpen
  \bibfield  {author} {\bibinfo {author} {\bibfnamefont {P.~J.}\ \bibnamefont
  {{McMillan}}},\ }\bibfield  {title} {\bibinfo {title} {{The mass distribution
  and gravitational potential of the Milky Way}},\ }\href
  {https://doi.org/10.1093/mnras/stw2759} {\bibfield  {journal} {\bibinfo
  {journal} {Mon. Not. Roy. Astron. Soc.}\ }\textbf {\bibinfo {volume} {465}},\
  \bibinfo {pages} {76} (\bibinfo {year} {2017})}\BibitemShut {NoStop}%
\bibitem [{Note1()}]{Note1}%
  \BibitemOpen
  \bibinfo {note} {In what follows, we assume that the spin of the pulsar's
  \protect \leavevmode {\protect \color {black}unseen companion neutron star
  (${\protect \bf S}_{\protect \rm c}$)} is negligible, i.e., $|{\protect \bf
  S_{\protect \rm c}|} \ll |{\protect \bf S_{\protect \rm p}}|$ so that
  ${\protect \bf J} \approx {\protect \bf L} + {\protect \bf S_{\protect \rm
  p}}$. \protect \leavevmode {\protect \color {black}This assumption is made
  based on expectations from binary stellar evolution, where the progenitor of
  the second-formed neutron star has donated mass onto the pulsar and therefore
  lost angular momentum prior to its supernova explosion.} LT precession in DNS
  systems will therefore \protect \leavevmode {\protect \color {black}most
  likely} be dominated by the spin of the observed pulsar.}\BibitemShut {Stop}%
\bibitem [{\citenamefont {{Wex}}\ and\ \citenamefont
  {{Kopeikin}}(1999)}]{wk99}%
  \BibitemOpen
  \bibfield  {author} {\bibinfo {author} {\bibfnamefont {N.}~\bibnamefont
  {{Wex}}}\ and\ \bibinfo {author} {\bibfnamefont {S.~M.}\ \bibnamefont
  {{Kopeikin}}},\ }\bibfield  {title} {\bibinfo {title} {{Frame Dragging and
  Other Precessional Effects in Black Hole Pulsar Binaries}},\ }\href
  {https://doi.org/10.1086/306933} {\bibfield  {journal} {\bibinfo  {journal}
  {Astrophys. J.}\ }\textbf {\bibinfo {volume} {514}},\ \bibinfo {pages} {388}
  (\bibinfo {year} {1999})}\BibitemShut {NoStop}%
\bibitem [{\citenamefont {{Damour}}\ and\ \citenamefont
  {{Taylor}}(1992)}]{dt92}%
  \BibitemOpen
  \bibfield  {author} {\bibinfo {author} {\bibfnamefont {T.}~\bibnamefont
  {{Damour}}}\ and\ \bibinfo {author} {\bibfnamefont {J.~H.}\ \bibnamefont
  {{Taylor}}},\ }\bibfield  {title} {\bibinfo {title} {{Strong-field tests of
  relativistic gravity and binary pulsars}},\ }\href
  {https://doi.org/10.1103/PhysRevD.45.1840} {\bibfield  {journal} {\bibinfo
  {journal} {Phys. Rev. D}\ }\textbf {\bibinfo {volume} {45}},\ \bibinfo
  {pages} {1840} (\bibinfo {year} {1992})}\BibitemShut {NoStop}%
\bibitem [{\citenamefont {{Damour}}\ and\ \citenamefont
  {{Taylor}}(1991)}]{dt91}%
  \BibitemOpen
  \bibfield  {author} {\bibinfo {author} {\bibfnamefont {T.}~\bibnamefont
  {{Damour}}}\ and\ \bibinfo {author} {\bibfnamefont {J.~H.}\ \bibnamefont
  {{Taylor}}},\ }\bibfield  {title} {\bibinfo {title} {{On the Orbital Period
  Change of the Binary Pulsar PSR 1913+16}},\ }\href
  {https://doi.org/10.1086/169585} {\bibfield  {journal} {\bibinfo  {journal}
  {Astrophys. J.}\ }\textbf {\bibinfo {volume} {366}},\ \bibinfo {pages} {501}
  (\bibinfo {year} {1991})}\BibitemShut {NoStop}%
\bibitem [{\citenamefont {{Peters}}(1964)}]{pet64}%
  \BibitemOpen
  \bibfield  {author} {\bibinfo {author} {\bibfnamefont {P.~C.}\ \bibnamefont
  {{Peters}}},\ }\bibfield  {title} {\bibinfo {title} {{Gravitational Radiation
  and the Motion of Two Point Masses}},\ }\href
  {https://doi.org/10.1103/PhysRev.136.B1224} {\bibfield  {journal} {\bibinfo
  {journal} {Phys. Rev.}\ }\textbf {\bibinfo {volume} {136}},\ \bibinfo {pages}
  {1224} (\bibinfo {year} {1964})}\BibitemShut {NoStop}%
\bibitem [{\citenamefont {{Kopeikin}}(1996)}]{kop96}%
  \BibitemOpen
  \bibfield  {author} {\bibinfo {author} {\bibfnamefont {S.~M.}\ \bibnamefont
  {{Kopeikin}}},\ }\bibfield  {title} {\bibinfo {title} {{Proper Motion of
  Binary Pulsars as a Source of Secular Variations of Orbital Parameters}},\
  }\href {https://doi.org/10.1086/310201} {\bibfield  {journal} {\bibinfo
  {journal} {Astrophys. J. Lett.}\ }\textbf {\bibinfo {volume} {467}},\
  \bibinfo {pages} {L93} (\bibinfo {year} {1996})}\BibitemShut {NoStop}%
\bibitem [{\citenamefont {{Barker}}\ and\ \citenamefont
  {{O'Connell}}(1975)}]{bo75}%
  \BibitemOpen
  \bibfield  {author} {\bibinfo {author} {\bibfnamefont {B.~M.}\ \bibnamefont
  {{Barker}}}\ and\ \bibinfo {author} {\bibfnamefont {R.~F.}\ \bibnamefont
  {{O'Connell}}},\ }\bibfield  {title} {\bibinfo {title} {{Gravitational
  two-body problem with arbitrary masses, spins, and quadrupole moments}},\
  }\href {https://doi.org/10.1103/PhysRevD.12.329} {\bibfield  {journal}
  {\bibinfo  {journal} {Phys. Rev. D}\ }\textbf {\bibinfo {volume} {12}},\
  \bibinfo {pages} {329} (\bibinfo {year} {1975})}\BibitemShut {NoStop}%
\bibitem [{Note2()}]{Note2}%
  \BibitemOpen
  \bibinfo {note} {Pulsar timing is only sensitive to radial displacements of
  the pulsar, which do not depend on \protect \leavevmode {\protect \color
  {black}the longitude of ascending node ($\Omega $) that is also shown in
  Figure \ref {fig:angles}.} The rate of change in $\Omega $ due to LT
  precession is therefore not directly measurable and is subsequently ignored
  in further discussion.}\BibitemShut {Stop}%
\bibitem [{\citenamefont {{Damour}}\ and\ \citenamefont
  {{Deruelle}}(1986)}]{dd86}%
  \BibitemOpen
  \bibfield  {author} {\bibinfo {author} {\bibfnamefont {T.}~\bibnamefont
  {{Damour}}}\ and\ \bibinfo {author} {\bibfnamefont {N.}~\bibnamefont
  {{Deruelle}}},\ }\bibfield  {title} {\bibinfo {title} {{General relativistic
  celestial mechanics of binary systems. II. The post-Newtonian timing
  formula.}},\ }\href@noop {} {\bibfield  {journal} {\bibinfo  {journal} {Ann.
  de L'Inst. Henri Poincare Sec. (A) Phys. Theor.}\ }\textbf {\bibinfo {volume}
  {44}},\ \bibinfo {pages} {263} (\bibinfo {year} {1986})}\BibitemShut
  {NoStop}%
\bibitem [{\citenamefont {{Luo}}\ \emph {et~al.}(2021)\citenamefont {{Luo}},
  \citenamefont {{Ransom}}, \citenamefont {{Demorest}}, \citenamefont {{Ray}},
  \citenamefont {{Archibald}}, \citenamefont {{Kerr}}, \citenamefont
  {{Jennings}}, \citenamefont {{Bachetti}}, \citenamefont {{van Haasteren}},
  \citenamefont {{Champagne}}, \citenamefont {{Colen}}, \citenamefont
  {{Phillips}}, \citenamefont {{Zimmerman}}, \citenamefont {{Stovall}},
  \citenamefont {{Lam}},\ and\ \citenamefont {{Jenet}}}]{lsd+21}%
  \BibitemOpen
  \bibfield  {author} {\bibinfo {author} {\bibfnamefont {J.}~\bibnamefont
  {{Luo}}}, \bibinfo {author} {\bibfnamefont {S.}~\bibnamefont {{Ransom}}},
  \bibinfo {author} {\bibfnamefont {P.}~\bibnamefont {{Demorest}}}, \bibinfo
  {author} {\bibfnamefont {P.~S.}\ \bibnamefont {{Ray}}}, \bibinfo {author}
  {\bibfnamefont {A.}~\bibnamefont {{Archibald}}}, \bibinfo {author}
  {\bibfnamefont {M.}~\bibnamefont {{Kerr}}}, \bibinfo {author} {\bibfnamefont
  {R.~J.}\ \bibnamefont {{Jennings}}}, \bibinfo {author} {\bibfnamefont
  {M.}~\bibnamefont {{Bachetti}}}, \bibinfo {author} {\bibfnamefont
  {R.}~\bibnamefont {{van Haasteren}}}, \bibinfo {author} {\bibfnamefont
  {C.~A.}\ \bibnamefont {{Champagne}}}, \bibinfo {author} {\bibfnamefont
  {J.}~\bibnamefont {{Colen}}}, \bibinfo {author} {\bibfnamefont
  {C.}~\bibnamefont {{Phillips}}}, \bibinfo {author} {\bibfnamefont
  {J.}~\bibnamefont {{Zimmerman}}}, \bibinfo {author} {\bibfnamefont
  {K.}~\bibnamefont {{Stovall}}}, \bibinfo {author} {\bibfnamefont {M.~T.}\
  \bibnamefont {{Lam}}},\ and\ \bibinfo {author} {\bibfnamefont {F.~A.}\
  \bibnamefont {{Jenet}}},\ }\bibfield  {title} {\bibinfo {title} {{PINT: A
  Modern Software Package for Pulsar Timing}},\ }\href
  {https://doi.org/10.3847/1538-4357/abe62f} {\bibfield  {journal} {\bibinfo
  {journal} {Astrophys. J.}\ }\textbf {\bibinfo {volume} {911}},\ \bibinfo
  {eid} {45} (\bibinfo {year} {2021})}\BibitemShut {NoStop}%
\bibitem [{\citenamefont {{Damour}}\ and\ \citenamefont
  {{Deruelle}}(1985)}]{dd85}%
  \BibitemOpen
  \bibfield  {author} {\bibinfo {author} {\bibfnamefont {T.}~\bibnamefont
  {{Damour}}}\ and\ \bibinfo {author} {\bibfnamefont {N.}~\bibnamefont
  {{Deruelle}}},\ }\bibfield  {title} {\bibinfo {title} {{General relativistic
  celestial mechanics of binary systems. I. The post-Newtonian motion.}},\
  }\href@noop {} {\bibfield  {journal} {\bibinfo  {journal} {Ann. de L'Inst.
  Henri Poincare Sec. (A) Phys. Theor.}\ }\textbf {\bibinfo {volume} {43}},\
  \bibinfo {pages} {107} (\bibinfo {year} {1985})}\BibitemShut {NoStop}%
\bibitem [{Note3()}]{Note3}%
  \BibitemOpen
  \bibinfo {note} {The value of $a_{\protect \rm p}$ was chosen to ensure that
  $\protect \qopname \relax o{cot}i \sim 0.1$ as required by the Keplerian mass
  function.}\BibitemShut {Stop}%
\bibitem [{\citenamefont {{Kramer}}\ \emph {et~al.}(2021)\citenamefont
  {{Kramer}}, \citenamefont {{Stairs}}, \citenamefont {{Manchester}},
  \citenamefont {{Wex}}, \citenamefont {{Deller}}, \citenamefont {{Coles}},
  \citenamefont {{Ali}}, \citenamefont {{Burgay}}, \citenamefont {{Camilo}},
  \citenamefont {{Cognard}}, \citenamefont {{Damour}}, \citenamefont
  {{Desvignes}}, \citenamefont {{Ferdman}}, \citenamefont {{Freire}},
  \citenamefont {{Grondin}}, \citenamefont {{Guillemot}}, \citenamefont
  {{Hobbs}}, \citenamefont {{Janssen}}, \citenamefont {{Karuppusamy}},
  \citenamefont {{Lorimer}}, \citenamefont {{Lyne}}, \citenamefont {{McKee}},
  \citenamefont {{McLaughlin}}, \citenamefont {{M{\"u}nch}}, \citenamefont
  {{Perera}}, \citenamefont {{Pol}}, \citenamefont {{Possenti}}, \citenamefont
  {{Sarkissian}}, \citenamefont {{Stappers}},\ and\ \citenamefont
  {{Theureau}}}]{ksm+21}%
  \BibitemOpen
  \bibfield  {author} {\bibinfo {author} {\bibfnamefont {M.}~\bibnamefont
  {{Kramer}}}, \bibinfo {author} {\bibfnamefont {I.~H.}\ \bibnamefont
  {{Stairs}}}, \bibinfo {author} {\bibfnamefont {R.~N.}\ \bibnamefont
  {{Manchester}}}, \bibinfo {author} {\bibfnamefont {N.}~\bibnamefont {{Wex}}},
  \bibinfo {author} {\bibfnamefont {A.~T.}\ \bibnamefont {{Deller}}}, \bibinfo
  {author} {\bibfnamefont {W.~A.}\ \bibnamefont {{Coles}}}, \bibinfo {author}
  {\bibfnamefont {M.}~\bibnamefont {{Ali}}}, \bibinfo {author} {\bibfnamefont
  {M.}~\bibnamefont {{Burgay}}}, \bibinfo {author} {\bibfnamefont
  {F.}~\bibnamefont {{Camilo}}}, \bibinfo {author} {\bibfnamefont
  {I.}~\bibnamefont {{Cognard}}}, \bibinfo {author} {\bibfnamefont
  {T.}~\bibnamefont {{Damour}}}, \bibinfo {author} {\bibfnamefont
  {G.}~\bibnamefont {{Desvignes}}}, \bibinfo {author} {\bibfnamefont {R.~D.}\
  \bibnamefont {{Ferdman}}}, \bibinfo {author} {\bibfnamefont {P.~C.~C.}\
  \bibnamefont {{Freire}}}, \bibinfo {author} {\bibfnamefont {S.}~\bibnamefont
  {{Grondin}}}, \bibinfo {author} {\bibfnamefont {L.}~\bibnamefont
  {{Guillemot}}}, \bibinfo {author} {\bibfnamefont {G.~B.}\ \bibnamefont
  {{Hobbs}}}, \bibinfo {author} {\bibfnamefont {G.}~\bibnamefont {{Janssen}}},
  \bibinfo {author} {\bibfnamefont {R.}~\bibnamefont {{Karuppusamy}}}, \bibinfo
  {author} {\bibfnamefont {D.~R.}\ \bibnamefont {{Lorimer}}}, \bibinfo {author}
  {\bibfnamefont {A.~G.}\ \bibnamefont {{Lyne}}}, \bibinfo {author}
  {\bibfnamefont {J.~W.}\ \bibnamefont {{McKee}}}, \bibinfo {author}
  {\bibfnamefont {M.}~\bibnamefont {{McLaughlin}}}, \bibinfo {author}
  {\bibfnamefont {L.~E.}\ \bibnamefont {{M{\"u}nch}}}, \bibinfo {author}
  {\bibfnamefont {B.~B.~P.}\ \bibnamefont {{Perera}}}, \bibinfo {author}
  {\bibfnamefont {N.}~\bibnamefont {{Pol}}}, \bibinfo {author} {\bibfnamefont
  {A.}~\bibnamefont {{Possenti}}}, \bibinfo {author} {\bibfnamefont
  {J.}~\bibnamefont {{Sarkissian}}}, \bibinfo {author} {\bibfnamefont {B.~W.}\
  \bibnamefont {{Stappers}}},\ and\ \bibinfo {author} {\bibfnamefont
  {G.}~\bibnamefont {{Theureau}}},\ }\bibfield  {title} {\bibinfo {title}
  {{Strong-Field Gravity Tests with the Double Pulsar}},\ }\href
  {https://doi.org/10.1103/PhysRevX.11.041050} {\bibfield  {journal} {\bibinfo
  {journal} {Phys. Rev. X}\ }\textbf {\bibinfo {volume} {11}},\ \bibinfo {eid}
  {041050} (\bibinfo {year} {2021})}\BibitemShut {NoStop}%
\bibitem [{\citenamefont {{Hu}}\ \emph {et~al.}(2022)\citenamefont {{Hu}},
  \citenamefont {{Kramer}}, \citenamefont {{Champion}}, \citenamefont {{Wex}},
  \citenamefont {{Parthasarathy}}, \citenamefont {{Pennucci}}, \citenamefont
  {{Porayko}}, \citenamefont {{van Straten}}, \citenamefont {{Venkatraman
  Krishnan}}, \citenamefont {{Burgay}}, \citenamefont {{Freire}}, \citenamefont
  {{Manchester}}, \citenamefont {{Possenti}}, \citenamefont {{Stairs}},
  \citenamefont {{Bailes}}, \citenamefont {{Buchner}}, \citenamefont
  {{Cameron}}, \citenamefont {{Camilo}},\ and\ \citenamefont
  {{Serylak}}}]{hkc+22}%
  \BibitemOpen
  \bibfield  {author} {\bibinfo {author} {\bibfnamefont {H.}~\bibnamefont
  {{Hu}}}, \bibinfo {author} {\bibfnamefont {M.}~\bibnamefont {{Kramer}}},
  \bibinfo {author} {\bibfnamefont {D.~J.}\ \bibnamefont {{Champion}}},
  \bibinfo {author} {\bibfnamefont {N.}~\bibnamefont {{Wex}}}, \bibinfo
  {author} {\bibfnamefont {A.}~\bibnamefont {{Parthasarathy}}}, \bibinfo
  {author} {\bibfnamefont {T.~T.}\ \bibnamefont {{Pennucci}}}, \bibinfo
  {author} {\bibfnamefont {N.~K.}\ \bibnamefont {{Porayko}}}, \bibinfo {author}
  {\bibfnamefont {W.}~\bibnamefont {{van Straten}}}, \bibinfo {author}
  {\bibfnamefont {V.}~\bibnamefont {{Venkatraman Krishnan}}}, \bibinfo {author}
  {\bibfnamefont {M.}~\bibnamefont {{Burgay}}}, \bibinfo {author}
  {\bibfnamefont {P.~C.~C.}\ \bibnamefont {{Freire}}}, \bibinfo {author}
  {\bibfnamefont {R.~N.}\ \bibnamefont {{Manchester}}}, \bibinfo {author}
  {\bibfnamefont {A.}~\bibnamefont {{Possenti}}}, \bibinfo {author}
  {\bibfnamefont {I.~H.}\ \bibnamefont {{Stairs}}}, \bibinfo {author}
  {\bibfnamefont {M.}~\bibnamefont {{Bailes}}}, \bibinfo {author}
  {\bibfnamefont {S.}~\bibnamefont {{Buchner}}}, \bibinfo {author}
  {\bibfnamefont {A.~D.}\ \bibnamefont {{Cameron}}}, \bibinfo {author}
  {\bibfnamefont {F.}~\bibnamefont {{Camilo}}},\ and\ \bibinfo {author}
  {\bibfnamefont {M.}~\bibnamefont {{Serylak}}},\ }\bibfield  {title} {\bibinfo
  {title} {{Gravitational signal propagation in the double pulsar studied with
  the MeerKAT telescope}},\ }\href
  {https://doi.org/10.1051/0004-6361/202244825} {\bibfield  {journal} {\bibinfo
   {journal} {Astro. \& Astrop.}\ }\textbf {\bibinfo {volume} {667}},\ \bibinfo
  {eid} {A149} (\bibinfo {year} {2022})}\BibitemShut {NoStop}%
\bibitem [{\citenamefont {{Stairs}}\ \emph {et~al.}(2004)\citenamefont
  {{Stairs}}, \citenamefont {{Thorsett}},\ and\ \citenamefont
  {{Arzoumanian}}}]{sta04}%
  \BibitemOpen
  \bibfield  {author} {\bibinfo {author} {\bibfnamefont {I.~H.}\ \bibnamefont
  {{Stairs}}}, \bibinfo {author} {\bibfnamefont {S.~E.}\ \bibnamefont
  {{Thorsett}}},\ and\ \bibinfo {author} {\bibfnamefont {Z.}~\bibnamefont
  {{Arzoumanian}}},\ }\bibfield  {title} {\bibinfo {title} {{Measurement of
  Gravitational Spin-Orbit Coupling in a Binary-Pulsar System}},\ }\href
  {https://doi.org/10.1103/PhysRevLett.93.141101} {\bibfield  {journal}
  {\bibinfo  {journal} {Phys. Rev. Lett.}\ }\textbf {\bibinfo {volume} {93}},\
  \bibinfo {eid} {141101} (\bibinfo {year} {2004})}\BibitemShut {NoStop}%
\bibitem [{\citenamefont {{Fonseca}}\ \emph {et~al.}(2014)\citenamefont
  {{Fonseca}}, \citenamefont {{Stairs}},\ and\ \citenamefont
  {{Thorsett}}}]{fst10}%
  \BibitemOpen
  \bibfield  {author} {\bibinfo {author} {\bibfnamefont {E.}~\bibnamefont
  {{Fonseca}}}, \bibinfo {author} {\bibfnamefont {I.~H.}\ \bibnamefont
  {{Stairs}}},\ and\ \bibinfo {author} {\bibfnamefont {S.~E.}\ \bibnamefont
  {{Thorsett}}},\ }\bibfield  {title} {\bibinfo {title} {{A Comprehensive Study
  of Relativistic Gravity Using PSR B1534+12}},\ }\href
  {https://doi.org/10.1088/0004-637X/787/1/82} {\bibfield  {journal} {\bibinfo
  {journal} {Astrophys. J.}\ }\textbf {\bibinfo {volume} {787}},\ \bibinfo
  {eid} {82} (\bibinfo {year} {2014})}\BibitemShut {NoStop}%
\bibitem [{\citenamefont {{Cameron}}\ \emph {et~al.}(2023)\citenamefont
  {{Cameron}}, \citenamefont {{Bailes}}, \citenamefont {{Champion}},
  \citenamefont {{Freire}}, \citenamefont {{Kramer}}, \citenamefont
  {{McLaughlin}}, \citenamefont {{Ng}}, \citenamefont {{Possenti}},
  \citenamefont {{Ridolfi}}, \citenamefont {{Tauris}}, \citenamefont {{Wahl}},\
  and\ \citenamefont {{Wex}}}]{cbc+23}%
  \BibitemOpen
  \bibfield  {author} {\bibinfo {author} {\bibfnamefont {A.~D.}\ \bibnamefont
  {{Cameron}}}, \bibinfo {author} {\bibfnamefont {M.}~\bibnamefont {{Bailes}}},
  \bibinfo {author} {\bibfnamefont {D.~J.}\ \bibnamefont {{Champion}}},
  \bibinfo {author} {\bibfnamefont {P.~C.~C.}\ \bibnamefont {{Freire}}},
  \bibinfo {author} {\bibfnamefont {M.}~\bibnamefont {{Kramer}}}, \bibinfo
  {author} {\bibfnamefont {M.~A.}\ \bibnamefont {{McLaughlin}}}, \bibinfo
  {author} {\bibfnamefont {C.}~\bibnamefont {{Ng}}}, \bibinfo {author}
  {\bibfnamefont {A.}~\bibnamefont {{Possenti}}}, \bibinfo {author}
  {\bibfnamefont {A.}~\bibnamefont {{Ridolfi}}}, \bibinfo {author}
  {\bibfnamefont {T.~M.}\ \bibnamefont {{Tauris}}}, \bibinfo {author}
  {\bibfnamefont {H.~M.}\ \bibnamefont {{Wahl}}},\ and\ \bibinfo {author}
  {\bibfnamefont {N.}~\bibnamefont {{Wex}}},\ }\bibfield  {title} {\bibinfo
  {title} {{New constraints on the kinematic, relativistic, and evolutionary
  properties of the PSR J1757-1854 double neutron star system}},\ }\href
  {https://doi.org/10.1093/mnras/stad1712} {\bibfield  {journal} {\bibinfo
  {journal} {Mon. Not. Roy. Astron. Soc.}\ }\textbf {\bibinfo {volume} {523}},\
  \bibinfo {pages} {5064} (\bibinfo {year} {2023})}\BibitemShut {NoStop}%
\bibitem [{\citenamefont {{Kramer}}(1998)}]{kra98}%
  \BibitemOpen
  \bibfield  {author} {\bibinfo {author} {\bibfnamefont {M.}~\bibnamefont
  {{Kramer}}},\ }\bibfield  {title} {\bibinfo {title} {{Determination of the
  Geometry of the PSR B1913+16 System by Geodetic Precession}},\ }\href
  {https://doi.org/10.1086/306535} {\bibfield  {journal} {\bibinfo  {journal}
  {Astrophys. J.}\ }\textbf {\bibinfo {volume} {509}},\ \bibinfo {pages} {856}
  (\bibinfo {year} {1998})}\BibitemShut {NoStop}%
\bibitem [{\citenamefont {{Weisberg}}\ and\ \citenamefont
  {{Huang}}(2016)}]{wh16}%
  \BibitemOpen
  \bibfield  {author} {\bibinfo {author} {\bibfnamefont {J.~M.}\ \bibnamefont
  {{Weisberg}}}\ and\ \bibinfo {author} {\bibfnamefont {Y.}~\bibnamefont
  {{Huang}}},\ }\bibfield  {title} {\bibinfo {title} {{Relativistic
  Measurements from Timing the Binary Pulsar PSR B1913+16}},\ }\href
  {https://doi.org/10.3847/0004-637X/829/1/55} {\bibfield  {journal} {\bibinfo
  {journal} {Astrophys. J.}\ }\textbf {\bibinfo {volume} {829}},\ \bibinfo
  {eid} {55} (\bibinfo {year} {2016})}\BibitemShut {NoStop}%
\bibitem [{\citenamefont {{Meng}}\ \emph {et~al.}(2024)\citenamefont {{Meng}},
  \citenamefont {{Zhu}}, \citenamefont {{Kramer}}, \citenamefont {{Miao}},
  \citenamefont {{Desvignes}}, \citenamefont {{Shao}}, \citenamefont {{Hu}},
  \citenamefont {{Freire}}, \citenamefont {{Zhang}}, \citenamefont {{Xue}},
  \citenamefont {{Fang}}, \citenamefont {{Champion}}, \citenamefont {{Yuan}},
  \citenamefont {{Miao}}, \citenamefont {{Niu}}, \citenamefont {{Fu}},
  \citenamefont {{Yao}}, \citenamefont {{Guo}},\ and\ \citenamefont
  {{Zhang}}}]{mzk+24}%
  \BibitemOpen
  \bibfield  {author} {\bibinfo {author} {\bibfnamefont {L.}~\bibnamefont
  {{Meng}}}, \bibinfo {author} {\bibfnamefont {W.}~\bibnamefont {{Zhu}}},
  \bibinfo {author} {\bibfnamefont {M.}~\bibnamefont {{Kramer}}}, \bibinfo
  {author} {\bibfnamefont {X.}~\bibnamefont {{Miao}}}, \bibinfo {author}
  {\bibfnamefont {G.}~\bibnamefont {{Desvignes}}}, \bibinfo {author}
  {\bibfnamefont {L.}~\bibnamefont {{Shao}}}, \bibinfo {author} {\bibfnamefont
  {H.}~\bibnamefont {{Hu}}}, \bibinfo {author} {\bibfnamefont {P.~C.~C.}\
  \bibnamefont {{Freire}}}, \bibinfo {author} {\bibfnamefont {Y.}~\bibnamefont
  {{Zhang}}}, \bibinfo {author} {\bibfnamefont {M.}~\bibnamefont {{Xue}}},
  \bibinfo {author} {\bibfnamefont {Z.}~\bibnamefont {{Fang}}}, \bibinfo
  {author} {\bibfnamefont {D.~J.}\ \bibnamefont {{Champion}}}, \bibinfo
  {author} {\bibfnamefont {M.}~\bibnamefont {{Yuan}}}, \bibinfo {author}
  {\bibfnamefont {C.}~\bibnamefont {{Miao}}}, \bibinfo {author} {\bibfnamefont
  {J.}~\bibnamefont {{Niu}}}, \bibinfo {author} {\bibfnamefont
  {Q.}~\bibnamefont {{Fu}}}, \bibinfo {author} {\bibfnamefont {J.}~\bibnamefont
  {{Yao}}}, \bibinfo {author} {\bibfnamefont {Y.}~\bibnamefont {{Guo}}},\ and\
  \bibinfo {author} {\bibfnamefont {C.}~\bibnamefont {{Zhang}}},\ }\bibfield
  {title} {\bibinfo {title} {{The Relativistic Spin Precession in the Compact
  Double Neutron Star System PSR J1946+2052}},\ }\href
  {https://doi.org/10.3847/1538-4357/ad381c} {\bibfield  {journal} {\bibinfo
  {journal} {Astrophys. J.}\ }\textbf {\bibinfo {volume} {966}},\ \bibinfo
  {eid} {46} (\bibinfo {year} {2024})}\BibitemShut {NoStop}%
\bibitem [{\citenamefont {{Meng}}\ \emph {et~al.}(2025)\citenamefont {{Meng}},
  \citenamefont {{Freire}}, \citenamefont {{Stovall}}, \citenamefont {{Wex}},
  \citenamefont {{Miao}}, \citenamefont {{Zhu}}, \citenamefont {{Kramer}},
  \citenamefont {{Cordes}}, \citenamefont {{Hu}}, \citenamefont {{Jiang}},
  \citenamefont {{Parent}}, \citenamefont {{Shao}}, \citenamefont {{Stairs}},
  \citenamefont {{Xue}}, \citenamefont {{Brazier}}, \citenamefont {{Camilo}},
  \citenamefont {{Champion}}, \citenamefont {{Chatterjee}}, \citenamefont
  {{Crawford}}, \citenamefont {{Fang}}, \citenamefont {{Fu}}, \citenamefont
  {{Guo}}, \citenamefont {{Hessels}}, \citenamefont {{MacLaughlin}},
  \citenamefont {{Miao}}, \citenamefont {{Niu}}, \citenamefont {{Wu}},
  \citenamefont {{Yao}}, \citenamefont {{Yuan}}, \citenamefont {{Yue}},\ and\
  \citenamefont {{Zhang}}}]{mfs+25}%
  \BibitemOpen
  \bibfield  {author} {\bibinfo {author} {\bibfnamefont {L.}~\bibnamefont
  {{Meng}}}, \bibinfo {author} {\bibfnamefont {P.~C.~C.}\ \bibnamefont
  {{Freire}}}, \bibinfo {author} {\bibfnamefont {K.}~\bibnamefont {{Stovall}}},
  \bibinfo {author} {\bibfnamefont {N.}~\bibnamefont {{Wex}}}, \bibinfo
  {author} {\bibfnamefont {X.}~\bibnamefont {{Miao}}}, \bibinfo {author}
  {\bibfnamefont {W.}~\bibnamefont {{Zhu}}}, \bibinfo {author} {\bibfnamefont
  {M.}~\bibnamefont {{Kramer}}}, \bibinfo {author} {\bibfnamefont {J.~M.}\
  \bibnamefont {{Cordes}}}, \bibinfo {author} {\bibfnamefont {H.}~\bibnamefont
  {{Hu}}}, \bibinfo {author} {\bibfnamefont {J.}~\bibnamefont {{Jiang}}},
  \bibinfo {author} {\bibfnamefont {E.}~\bibnamefont {{Parent}}}, \bibinfo
  {author} {\bibfnamefont {L.}~\bibnamefont {{Shao}}}, \bibinfo {author}
  {\bibfnamefont {I.~H.}\ \bibnamefont {{Stairs}}}, \bibinfo {author}
  {\bibfnamefont {M.}~\bibnamefont {{Xue}}}, \bibinfo {author} {\bibfnamefont
  {A.}~\bibnamefont {{Brazier}}}, \bibinfo {author} {\bibfnamefont
  {F.}~\bibnamefont {{Camilo}}}, \bibinfo {author} {\bibfnamefont {D.~J.}\
  \bibnamefont {{Champion}}}, \bibinfo {author} {\bibfnamefont
  {S.}~\bibnamefont {{Chatterjee}}}, \bibinfo {author} {\bibfnamefont
  {F.}~\bibnamefont {{Crawford}}}, \bibinfo {author} {\bibfnamefont
  {Z.}~\bibnamefont {{Fang}}}, \bibinfo {author} {\bibfnamefont
  {Q.}~\bibnamefont {{Fu}}}, \bibinfo {author} {\bibfnamefont {Y.}~\bibnamefont
  {{Guo}}}, \bibinfo {author} {\bibfnamefont {J.~W.~T.}\ \bibnamefont
  {{Hessels}}}, \bibinfo {author} {\bibfnamefont {M.}~\bibnamefont
  {{MacLaughlin}}}, \bibinfo {author} {\bibfnamefont {C.}~\bibnamefont
  {{Miao}}}, \bibinfo {author} {\bibfnamefont {J.}~\bibnamefont {{Niu}}},
  \bibinfo {author} {\bibfnamefont {Z.}~\bibnamefont {{Wu}}}, \bibinfo {author}
  {\bibfnamefont {J.}~\bibnamefont {{Yao}}}, \bibinfo {author} {\bibfnamefont
  {M.}~\bibnamefont {{Yuan}}}, \bibinfo {author} {\bibfnamefont
  {Y.}~\bibnamefont {{Yue}}},\ and\ \bibinfo {author} {\bibfnamefont
  {C.}~\bibnamefont {{Zhang}}},\ }\bibfield  {title} {\bibinfo {title} {{The
  double neutron star PSR J1946+2052 I. Masses and tests of general
  relativity}},\ }\href {https://doi.org/10.48550/arXiv.2510.12506} {\bibfield
  {journal} {\bibinfo  {journal} {arXiv e-prints}\ ,\ \bibinfo {eid}
  {arXiv:2510.12506}} (\bibinfo {year} {2025})}\BibitemShut {NoStop}%
\bibitem [{\citenamefont {{Hallinan}}\ \emph {et~al.}(2019)\citenamefont
  {{Hallinan}}, \citenamefont {{Ravi}}, \citenamefont {{Weinreb}},
  \citenamefont {{Kocz}}, \citenamefont {{Huang}}, \citenamefont {{Woody}},
  \citenamefont {{Lamb}}, \citenamefont {{D'Addario}}, \citenamefont {{Catha}},
  \citenamefont {{Law}}, \citenamefont {{Kulkarni}}, \citenamefont {{Phinney}},
  \citenamefont {{Eastwood}}, \citenamefont {{Bouman}}, \citenamefont
  {{McLaughlin}}, \citenamefont {{Ransom}}, \citenamefont {{Siemens}},
  \citenamefont {{Cordes}}, \citenamefont {{Lynch}}, \citenamefont {{Kaplan}},
  \citenamefont {{Brazier}}, \citenamefont {{Bhatnagar}}, \citenamefont
  {{Myers}}, \citenamefont {{Walter}},\ and\ \citenamefont
  {{Gaensler}}}]{hrw+19}%
  \BibitemOpen
  \bibfield  {author} {\bibinfo {author} {\bibfnamefont {G.}~\bibnamefont
  {{Hallinan}}}, \bibinfo {author} {\bibfnamefont {V.}~\bibnamefont {{Ravi}}},
  \bibinfo {author} {\bibfnamefont {S.}~\bibnamefont {{Weinreb}}}, \bibinfo
  {author} {\bibfnamefont {J.}~\bibnamefont {{Kocz}}}, \bibinfo {author}
  {\bibfnamefont {Y.}~\bibnamefont {{Huang}}}, \bibinfo {author} {\bibfnamefont
  {D.~P.}\ \bibnamefont {{Woody}}}, \bibinfo {author} {\bibfnamefont
  {J.}~\bibnamefont {{Lamb}}}, \bibinfo {author} {\bibfnamefont
  {L.}~\bibnamefont {{D'Addario}}}, \bibinfo {author} {\bibfnamefont
  {M.}~\bibnamefont {{Catha}}}, \bibinfo {author} {\bibfnamefont
  {C.}~\bibnamefont {{Law}}}, \bibinfo {author} {\bibfnamefont {S.~R.}\
  \bibnamefont {{Kulkarni}}}, \bibinfo {author} {\bibfnamefont {E.~S.}\
  \bibnamefont {{Phinney}}}, \bibinfo {author} {\bibfnamefont {M.~W.}\
  \bibnamefont {{Eastwood}}}, \bibinfo {author} {\bibfnamefont
  {K.}~\bibnamefont {{Bouman}}}, \bibinfo {author} {\bibfnamefont
  {M.}~\bibnamefont {{McLaughlin}}}, \bibinfo {author} {\bibfnamefont
  {S.}~\bibnamefont {{Ransom}}}, \bibinfo {author} {\bibfnamefont
  {X.}~\bibnamefont {{Siemens}}}, \bibinfo {author} {\bibfnamefont
  {J.}~\bibnamefont {{Cordes}}}, \bibinfo {author} {\bibfnamefont
  {R.}~\bibnamefont {{Lynch}}}, \bibinfo {author} {\bibfnamefont
  {D.}~\bibnamefont {{Kaplan}}}, \bibinfo {author} {\bibfnamefont
  {A.}~\bibnamefont {{Brazier}}}, \bibinfo {author} {\bibfnamefont
  {S.}~\bibnamefont {{Bhatnagar}}}, \bibinfo {author} {\bibfnamefont
  {S.}~\bibnamefont {{Myers}}}, \bibinfo {author} {\bibfnamefont
  {F.}~\bibnamefont {{Walter}}},\ and\ \bibinfo {author} {\bibfnamefont
  {B.}~\bibnamefont {{Gaensler}}},\ }\bibfield  {title} {\bibinfo {title} {{The
  DSA-2000 {\textemdash} A Radio Survey Camera}},\ }in\ \href
  {https://doi.org/10.48550/arXiv.1907.07648} {\emph {\bibinfo {booktitle}
  {Bulletin of the American Astronomical Society}}},\ Vol.~\bibinfo {volume}
  {51}\ (\bibinfo {year} {2019})\ p.\ \bibinfo {pages} {255}\BibitemShut
  {NoStop}%
\bibitem [{Note4()}]{Note4}%
  \BibitemOpen
  \bibinfo {note} {Each curve in these figures represents the average of 10
  simulations that each generate different realizations of statistically white
  noise with standard deviation equal to $\sigma _{\protect \rm RMS}$. This
  averaging is performed to minimize short-term fluctuations due to
  noise.}\BibitemShut {Stop}%
\bibitem [{\citenamefont {{Jones}}\ \emph {et~al.}(2017)\citenamefont
  {{Jones}}, \citenamefont {{McLaughlin}}, \citenamefont {{Lam}}, \citenamefont
  {{Cordes}}, \citenamefont {{Levin}}, \citenamefont {{Chatterjee}},
  \citenamefont {{Arzoumanian}}, \citenamefont {{Crowter}}, \citenamefont
  {{Demorest}}, \citenamefont {{Dolch}}, \citenamefont {{Ellis}}, \citenamefont
  {{Ferdman}}, \citenamefont {{Fonseca}}, \citenamefont {{Gonzalez}},
  \citenamefont {{Jones}}, \citenamefont {{Lazio}}, \citenamefont {{Nice}},
  \citenamefont {{Pennucci}}, \citenamefont {{Ransom}}, \citenamefont
  {{Stinebring}}, \citenamefont {{Stairs}}, \citenamefont {{Stovall}},
  \citenamefont {{Swiggum}},\ and\ \citenamefont {{Zhu}}}]{mml+17}%
  \BibitemOpen
  \bibfield  {author} {\bibinfo {author} {\bibfnamefont {M.~L.}\ \bibnamefont
  {{Jones}}}, \bibinfo {author} {\bibfnamefont {M.~A.}\ \bibnamefont
  {{McLaughlin}}}, \bibinfo {author} {\bibfnamefont {M.~T.}\ \bibnamefont
  {{Lam}}}, \bibinfo {author} {\bibfnamefont {J.~M.}\ \bibnamefont {{Cordes}}},
  \bibinfo {author} {\bibfnamefont {L.}~\bibnamefont {{Levin}}}, \bibinfo
  {author} {\bibfnamefont {S.}~\bibnamefont {{Chatterjee}}}, \bibinfo {author}
  {\bibfnamefont {Z.}~\bibnamefont {{Arzoumanian}}}, \bibinfo {author}
  {\bibfnamefont {K.}~\bibnamefont {{Crowter}}}, \bibinfo {author}
  {\bibfnamefont {P.~B.}\ \bibnamefont {{Demorest}}}, \bibinfo {author}
  {\bibfnamefont {T.}~\bibnamefont {{Dolch}}}, \bibinfo {author} {\bibfnamefont
  {J.~A.}\ \bibnamefont {{Ellis}}}, \bibinfo {author} {\bibfnamefont {R.~D.}\
  \bibnamefont {{Ferdman}}}, \bibinfo {author} {\bibfnamefont {E.}~\bibnamefont
  {{Fonseca}}}, \bibinfo {author} {\bibfnamefont {M.~E.}\ \bibnamefont
  {{Gonzalez}}}, \bibinfo {author} {\bibfnamefont {G.}~\bibnamefont {{Jones}}},
  \bibinfo {author} {\bibfnamefont {T.~J.~W.}\ \bibnamefont {{Lazio}}},
  \bibinfo {author} {\bibfnamefont {D.~J.}\ \bibnamefont {{Nice}}}, \bibinfo
  {author} {\bibfnamefont {T.~T.}\ \bibnamefont {{Pennucci}}}, \bibinfo
  {author} {\bibfnamefont {S.~M.}\ \bibnamefont {{Ransom}}}, \bibinfo {author}
  {\bibfnamefont {D.~R.}\ \bibnamefont {{Stinebring}}}, \bibinfo {author}
  {\bibfnamefont {I.~H.}\ \bibnamefont {{Stairs}}}, \bibinfo {author}
  {\bibfnamefont {K.}~\bibnamefont {{Stovall}}}, \bibinfo {author}
  {\bibfnamefont {J.~K.}\ \bibnamefont {{Swiggum}}},\ and\ \bibinfo {author}
  {\bibfnamefont {W.~W.}\ \bibnamefont {{Zhu}}},\ }\bibfield  {title} {\bibinfo
  {title} {{The NANOGrav Nine-year Data Set: Measurement and Analysis of
  Variations in Dispersion Measures}},\ }\href
  {https://doi.org/10.3847/1538-4357/aa73df} {\bibfield  {journal} {\bibinfo
  {journal} {Astrophys. J.}\ }\textbf {\bibinfo {volume} {841}},\ \bibinfo
  {eid} {125} (\bibinfo {year} {2017})}\BibitemShut {NoStop}%
\bibitem [{\citenamefont {{Antonelli}}\ \emph {et~al.}(2023)\citenamefont
  {{Antonelli}}, \citenamefont {{Basu}},\ and\ \citenamefont
  {{Haskell}}}]{abh23}%
  \BibitemOpen
  \bibfield  {author} {\bibinfo {author} {\bibfnamefont {M.}~\bibnamefont
  {{Antonelli}}}, \bibinfo {author} {\bibfnamefont {A.}~\bibnamefont
  {{Basu}}},\ and\ \bibinfo {author} {\bibfnamefont {B.}~\bibnamefont
  {{Haskell}}},\ }\bibfield  {title} {\bibinfo {title} {{Stochastic processes
  for pulsar timing noise: fluctuations in the internal and external
  torques}},\ }\href {https://doi.org/10.1093/mnras/stad256} {\bibfield
  {journal} {\bibinfo  {journal} {Mon. Not. Roy. Astron. Soc.}\ }\textbf
  {\bibinfo {volume} {520}},\ \bibinfo {pages} {2813} (\bibinfo {year}
  {2023})}\BibitemShut {NoStop}%
\bibitem [{\citenamefont {{Hu}}(2025)}]{hu25}%
  \BibitemOpen
  \bibfield  {author} {\bibinfo {author} {\bibfnamefont {H.}~\bibnamefont
  {{Hu}}},\ }\bibfield  {title} {\bibinfo {title} {{Unlocking gravity and
  gravitational waves with radio pulsars: advances and challenges}},\ }\href
  {https://doi.org/10.1007/s10509-025-04463-2} {\bibfield  {journal} {\bibinfo
  {journal} {Astrophys. Spa. Sci.}\ }\textbf {\bibinfo {volume} {370}},\
  \bibinfo {eid} {74} (\bibinfo {year} {2025})}\BibitemShut {NoStop}%
\bibitem [{\citenamefont {{Reardon}}\ \emph {et~al.}(2023)\citenamefont
  {{Reardon}}, \citenamefont {{Zic}}, \citenamefont {{Shannon}}, \citenamefont
  {{Di Marco}}, \citenamefont {{Hobbs}}, \citenamefont {{Kapur}}, \citenamefont
  {{Lower}}, \citenamefont {{Mandow}}, \citenamefont {{Middleton}},
  \citenamefont {{Miles}}, \citenamefont {{Rogers}}, \citenamefont {{Askew}},
  \citenamefont {{Bailes}}, \citenamefont {{Bhat}}, \citenamefont {{Cameron}},
  \citenamefont {{Kerr}}, \citenamefont {{Kulkarni}}, \citenamefont
  {{Manchester}}, \citenamefont {{Nathan}}, \citenamefont {{Russell}},
  \citenamefont {{Os{\l}owski}},\ and\ \citenamefont {{Zhu}}}]{rzs+23}%
  \BibitemOpen
  \bibfield  {author} {\bibinfo {author} {\bibfnamefont {D.~J.}\ \bibnamefont
  {{Reardon}}}, \bibinfo {author} {\bibfnamefont {A.}~\bibnamefont {{Zic}}},
  \bibinfo {author} {\bibfnamefont {R.~M.}\ \bibnamefont {{Shannon}}}, \bibinfo
  {author} {\bibfnamefont {V.}~\bibnamefont {{Di Marco}}}, \bibinfo {author}
  {\bibfnamefont {G.~B.}\ \bibnamefont {{Hobbs}}}, \bibinfo {author}
  {\bibfnamefont {A.}~\bibnamefont {{Kapur}}}, \bibinfo {author} {\bibfnamefont
  {M.~E.}\ \bibnamefont {{Lower}}}, \bibinfo {author} {\bibfnamefont
  {R.}~\bibnamefont {{Mandow}}}, \bibinfo {author} {\bibfnamefont
  {H.}~\bibnamefont {{Middleton}}}, \bibinfo {author} {\bibfnamefont {M.~T.}\
  \bibnamefont {{Miles}}}, \bibinfo {author} {\bibfnamefont {A.~F.}\
  \bibnamefont {{Rogers}}}, \bibinfo {author} {\bibfnamefont {J.}~\bibnamefont
  {{Askew}}}, \bibinfo {author} {\bibfnamefont {M.}~\bibnamefont {{Bailes}}},
  \bibinfo {author} {\bibfnamefont {N.~D.~R.}\ \bibnamefont {{Bhat}}}, \bibinfo
  {author} {\bibfnamefont {A.}~\bibnamefont {{Cameron}}}, \bibinfo {author}
  {\bibfnamefont {M.}~\bibnamefont {{Kerr}}}, \bibinfo {author} {\bibfnamefont
  {A.}~\bibnamefont {{Kulkarni}}}, \bibinfo {author} {\bibfnamefont {R.~N.}\
  \bibnamefont {{Manchester}}}, \bibinfo {author} {\bibfnamefont {R.~S.}\
  \bibnamefont {{Nathan}}}, \bibinfo {author} {\bibfnamefont {C.~J.}\
  \bibnamefont {{Russell}}}, \bibinfo {author} {\bibfnamefont {S.}~\bibnamefont
  {{Os{\l}owski}}},\ and\ \bibinfo {author} {\bibfnamefont {X.-J.}\
  \bibnamefont {{Zhu}}},\ }\bibfield  {title} {\bibinfo {title} {{The
  Gravitational-wave Background Null Hypothesis: Characterizing Noise in
  Millisecond Pulsar Arrival Times with the Parkes Pulsar Timing Array}},\
  }\href {https://doi.org/10.3847/2041-8213/acdd03} {\bibfield  {journal}
  {\bibinfo  {journal} {Astrophys. J. Lett.}\ }\textbf {\bibinfo {volume}
  {951}},\ \bibinfo {eid} {L7} (\bibinfo {year} {2023})}\BibitemShut {NoStop}%
\bibitem [{\citenamefont {{Bates}}\ \emph {et~al.}(2013)\citenamefont
  {{Bates}}, \citenamefont {{Lorimer}},\ and\ \citenamefont
  {{Verbiest}}}]{blv+13}%
  \BibitemOpen
  \bibfield  {author} {\bibinfo {author} {\bibfnamefont {S.~D.}\ \bibnamefont
  {{Bates}}}, \bibinfo {author} {\bibfnamefont {D.~R.}\ \bibnamefont
  {{Lorimer}}},\ and\ \bibinfo {author} {\bibfnamefont {J.~P.~W.}\ \bibnamefont
  {{Verbiest}}},\ }\bibfield  {title} {\bibinfo {title} {{The pulsar spectral
  index distribution}},\ }\href {https://doi.org/10.1093/mnras/stt257}
  {\bibfield  {journal} {\bibinfo  {journal} {Mon. Not. Roy. Astron. Soc.}\
  }\textbf {\bibinfo {volume} {431}},\ \bibinfo {pages} {1352} (\bibinfo {year}
  {2013})}\BibitemShut {NoStop}%
\bibitem [{\citenamefont {{Konacki}}\ \emph {et~al.}(2003)\citenamefont
  {{Konacki}}, \citenamefont {{Wolszczan}},\ and\ \citenamefont
  {{Stairs}}}]{kws03}%
  \BibitemOpen
  \bibfield  {author} {\bibinfo {author} {\bibfnamefont {M.}~\bibnamefont
  {{Konacki}}}, \bibinfo {author} {\bibfnamefont {A.}~\bibnamefont
  {{Wolszczan}}},\ and\ \bibinfo {author} {\bibfnamefont {I.~H.}\ \bibnamefont
  {{Stairs}}},\ }\bibfield  {title} {\bibinfo {title} {{Geodetic Precession and
  Timing of the Relativistic Binary Pulsars PSR B1534+12 and PSR B1913+16}},\
  }\href {https://doi.org/10.1086/374418} {\bibfield  {journal} {\bibinfo
  {journal} {Astrophys. J.}\ }\textbf {\bibinfo {volume} {589}},\ \bibinfo
  {pages} {495} (\bibinfo {year} {2003})}\BibitemShut {NoStop}%
\bibitem [{\citenamefont {{van Leeuwen}}\ \emph {et~al.}(2015)\citenamefont
  {{van Leeuwen}}, \citenamefont {{Kasian}}, \citenamefont {{Stairs}},
  \citenamefont {{Lorimer}}, \citenamefont {{Camilo}}, \citenamefont
  {{Chatterjee}}, \citenamefont {{Cognard}}, \citenamefont {{Desvignes}},
  \citenamefont {{Freire}}, \citenamefont {{Janssen}}, \citenamefont
  {{Kramer}}, \citenamefont {{Lyne}}, \citenamefont {{Nice}}, \citenamefont
  {{Ransom}}, \citenamefont {{Stappers}},\ and\ \citenamefont
  {{Weisberg}}}]{vks+15}%
  \BibitemOpen
  \bibfield  {author} {\bibinfo {author} {\bibfnamefont {J.}~\bibnamefont {{van
  Leeuwen}}}, \bibinfo {author} {\bibfnamefont {L.}~\bibnamefont {{Kasian}}},
  \bibinfo {author} {\bibfnamefont {I.~H.}\ \bibnamefont {{Stairs}}}, \bibinfo
  {author} {\bibfnamefont {D.~R.}\ \bibnamefont {{Lorimer}}}, \bibinfo {author}
  {\bibfnamefont {F.}~\bibnamefont {{Camilo}}}, \bibinfo {author}
  {\bibfnamefont {S.}~\bibnamefont {{Chatterjee}}}, \bibinfo {author}
  {\bibfnamefont {I.}~\bibnamefont {{Cognard}}}, \bibinfo {author}
  {\bibfnamefont {G.}~\bibnamefont {{Desvignes}}}, \bibinfo {author}
  {\bibfnamefont {P.~C.~C.}\ \bibnamefont {{Freire}}}, \bibinfo {author}
  {\bibfnamefont {G.~H.}\ \bibnamefont {{Janssen}}}, \bibinfo {author}
  {\bibfnamefont {M.}~\bibnamefont {{Kramer}}}, \bibinfo {author}
  {\bibfnamefont {A.~G.}\ \bibnamefont {{Lyne}}}, \bibinfo {author}
  {\bibfnamefont {D.~J.}\ \bibnamefont {{Nice}}}, \bibinfo {author}
  {\bibfnamefont {S.~M.}\ \bibnamefont {{Ransom}}}, \bibinfo {author}
  {\bibfnamefont {B.~W.}\ \bibnamefont {{Stappers}}},\ and\ \bibinfo {author}
  {\bibfnamefont {J.~M.}\ \bibnamefont {{Weisberg}}},\ }\bibfield  {title}
  {\bibinfo {title} {{The Binary Companion of Young, Relativistic Pulsar
  J1906+0746}},\ }\href {https://doi.org/10.1088/0004-637X/798/2/118}
  {\bibfield  {journal} {\bibinfo  {journal} {Astrophys. J.}\ }\textbf
  {\bibinfo {volume} {798}},\ \bibinfo {eid} {118} (\bibinfo {year}
  {2015})}\BibitemShut {NoStop}%
\bibitem [{\citenamefont {{Desvignes}}\ \emph {et~al.}(2019)\citenamefont
  {{Desvignes}}, \citenamefont {{Kramer}}, \citenamefont {{Lee}}, \citenamefont
  {{van Leeuwen}}, \citenamefont {{Stairs}}, \citenamefont {{Jessner}},
  \citenamefont {{Cognard}}, \citenamefont {{Kasian}}, \citenamefont {{Lyne}},\
  and\ \citenamefont {{Stappers}}}]{dkl+19}%
  \BibitemOpen
  \bibfield  {author} {\bibinfo {author} {\bibfnamefont {G.}~\bibnamefont
  {{Desvignes}}}, \bibinfo {author} {\bibfnamefont {M.}~\bibnamefont
  {{Kramer}}}, \bibinfo {author} {\bibfnamefont {K.}~\bibnamefont {{Lee}}},
  \bibinfo {author} {\bibfnamefont {J.}~\bibnamefont {{van Leeuwen}}}, \bibinfo
  {author} {\bibfnamefont {I.}~\bibnamefont {{Stairs}}}, \bibinfo {author}
  {\bibfnamefont {A.}~\bibnamefont {{Jessner}}}, \bibinfo {author}
  {\bibfnamefont {I.}~\bibnamefont {{Cognard}}}, \bibinfo {author}
  {\bibfnamefont {L.}~\bibnamefont {{Kasian}}}, \bibinfo {author}
  {\bibfnamefont {A.}~\bibnamefont {{Lyne}}},\ and\ \bibinfo {author}
  {\bibfnamefont {B.~W.}\ \bibnamefont {{Stappers}}},\ }\bibfield  {title}
  {\bibinfo {title} {{Radio emission from a pulsar{\textquoteright}s magnetic
  pole revealed by general relativity}},\ }\href
  {https://doi.org/10.1126/science.aav7272} {\bibfield  {journal} {\bibinfo
  {journal} {Science}\ }\textbf {\bibinfo {volume} {365}},\ \bibinfo {pages}
  {1013} (\bibinfo {year} {2019})}\BibitemShut {NoStop}%
\bibitem [{\citenamefont {{Vleeschower}}\ \emph {et~al.}(2026)\citenamefont
  {{Vleeschower}}, \citenamefont {{Stappers}}, \citenamefont {{Keith}},
  \citenamefont {{Desvignes}}, \citenamefont {{Freire}}, \citenamefont
  {{Kramer}}, \citenamefont {{van Leeuwen}}, \citenamefont {{Levin}},
  \citenamefont {{Lyne}}, \citenamefont {{Stairs}}, \citenamefont
  {{Krishnan}},\ and\ \citenamefont {{Wang}}}]{vsk+26}%
  \BibitemOpen
  \bibfield  {author} {\bibinfo {author} {\bibfnamefont {L.}~\bibnamefont
  {{Vleeschower}}}, \bibinfo {author} {\bibfnamefont {B.~W.}\ \bibnamefont
  {{Stappers}}}, \bibinfo {author} {\bibfnamefont {M.~J.}\ \bibnamefont
  {{Keith}}}, \bibinfo {author} {\bibfnamefont {G.}~\bibnamefont
  {{Desvignes}}}, \bibinfo {author} {\bibfnamefont {P.~C.~C.}\ \bibnamefont
  {{Freire}}}, \bibinfo {author} {\bibfnamefont {M.}~\bibnamefont {{Kramer}}},
  \bibinfo {author} {\bibfnamefont {J.}~\bibnamefont {{van Leeuwen}}}, \bibinfo
  {author} {\bibfnamefont {L.}~\bibnamefont {{Levin}}}, \bibinfo {author}
  {\bibfnamefont {A.~G.}\ \bibnamefont {{Lyne}}}, \bibinfo {author}
  {\bibfnamefont {I.~H.}\ \bibnamefont {{Stairs}}}, \bibinfo {author}
  {\bibfnamefont {V.~V.}\ \bibnamefont {{Krishnan}}},\ and\ \bibinfo {author}
  {\bibfnamefont {Y.~Y.}\ \bibnamefont {{Wang}}},\ }\bibfield  {title}
  {\bibinfo {title} {{Long-term timing of the relativistic binary PSR
  J1906+0746}},\ }\href {https://doi.org/10.1093/mnras/stag264} {\bibfield
  {journal} {\bibinfo  {journal} {MNRAS}\ }\textbf {\bibinfo {volume} {546}},\
  \bibinfo {eid} {stag264} (\bibinfo {year} {2026})}\BibitemShut {NoStop}%
\end{thebibliography}%

\end{document}